\documentclass[letterpaper, 10 pt, conference]{ieeeconf}  

\IEEEoverridecommandlockouts                              

\usepackage{cite}
\usepackage{graphics} 
\usepackage{epsfig} 
\usepackage{times} 
\usepackage{amsmath,bm}
\usepackage{amssymb}  
\usepackage{cases}

\DeclareMathOperator{\sign}{sign}

\usepackage{mathtools}

\usepackage{caption}
\usepackage{subcaption}

\usepackage[usenames, dvipsnames]{color}
\newtheorem{assumption}{Assumption}
\newtheorem{thm}{Theorem}
\newtheorem{lem}{Lemma}

\newtheorem{prob}{Problem}

\usepackage{graphicx}
\usepackage{float}
\usepackage{xcolor}

\usepackage{siunitx}

\usepackage{hyperref}
\hypersetup{
    colorlinks=true,
    linkcolor=black,
    citecolor=black,
    urlcolor=blue,
}

\title{\LARGE \bf
Safe platooning control of connected and autonomous vehicles on curved multi-lane roads  }

\author{Xiao Chen$^{1}$, Zhiqi Tang$^{1}$, Karl H. Johansson$^{1}$ and Jonas Mårtensson$^{1}$
\thanks{$^{1}$ Division of Decision and Control Systems, School of Electrical Engineering and Computer Science, KTH Royal Institute of Technology, Sweden. Emails: {\tt\small \{xiao2, ztang2, kallej, jonas1\}@kth.se}}%
}

\begin{document}

\maketitle
\thispagestyle{empty}
\pagestyle{empty}

\begin{abstract}

This paper investigates the safe platoon formation tracking and merging control problem of connected and automated vehicles (CAVs) on curved multi-lane  roads. The first novelty is the separation of the control designs into two distinct parts: a lateral control law that ensures a geometrical convergence towards the reference path regardless of the translational velocity, and a longitudinal control design for each vehicle
to achieve the desired relative arc length and velocity with
respect to its neighboring vehicle. The second novelty is exploiting the constructive barrier feedback as an additive term to the nominal tracking control, ensuring both lateral and longitudinal collision avoidance. This constructive barrier feedback acts as a dissipative term, slowing down the relative velocity toward obstacles without affecting the nominal controller's performance. Consequently, our proposed control method enables safe platoon formation of vehicles on curved multi-lane roads, with theoretical guarantees for safety invariance and stability analysis. Simulation and experimental results on connected vehicles are provided to further validate the effectiveness of the proposed method.
\end{abstract}


\section{Introduction}
Interest in connected and automated vehicles (CAVs) has surged in recent years, driven by advancements in automation and communication technologies. CAVs offer the potential for cooperative on-road operations, promising significant benefits to the transportation sector. Vehicle platooning has emerged as a popular concept among the various forms of cooperative driving technology. Platooning involves a train of CAVs operating with minimal inter-vehicle distances and synchronized speeds. The exploration of vehicle platooning traces back to the early 2000s, initiated by pioneering projects such as PATH \cite{Path} and Sartre \cite{Sartre}, which aimed to assess the potential benefits of this technology. Subsequent studies such as \cite{platoonbeneanalysis,platooningBenefit,turri2016cooperative} have reinforced that vehicle platooning is promising to improve traffic efficiency, enhance safety, and reduce energy consumption.

Earlier works on designing control techniques to support vehicle platooning mainly focused on the simplified one-dimensional longitudinal control of vehicles for distance keeping and velocity synchronization by guaranteeing string stability through the design \cite{platoonlinearstability, Slidingpractical, slidningintegral}. In a more general multi-lane traffic setting, muti-vehicle platooning requires adaptive formation adjustments of vehicles from different lanes while accommodating the road’s shape. The key challenge here is to design efficient platoon control strategies for multi-vehicle systems in structured road environments, preventing vehicle-to-vehicle collisions and encroachment on road boundaries. 

 Most of the existing work in the literature about safe platooning and merging control in multi-lane scenarios is restricted to the simplified type of roads with straight lines, and a majority of them typically focus only on lateral or longitude controller design. For instance, the work in \cite{lyapunov2Dcruise} adopts potential function-based control strategies for a group of vehicles autonomously driving on the straight-line road with velocity consensus. Some other work uses optimization-based controllers, such as model predictive control and barrier function-based optimization controllers. In the work of \cite{MPCdistributed}, a safety-assuring MPC framework is developed, however, focusing only on safe lateral merging maneuvers of vehicles on straight-line roads.  Another approach in \cite{CBFlongi}  exploits barrier function-based optimization controllers but only for longitudinal merging control. 
 It is worth noting that the use of optimization-based controllers poses challenges in explicitly
analyzing the equilibrium and convergence of the multi-vehicle
system, in addition to potential computational complexity and feasibility issues. To the best of our knowledge, the control problems of safe platooning and merging for CAVs on multi-laned and curved roads remain open.

Motivated by the above-mentioned open problems, in this paper, we propose a novel decentralized control strategy for the safe platooning and merging problem oncurved multi-lane roads, covering both longitudinal and lateral design. Each vehicle is modeled as a nonholonomic second-order kinematic bicycle model. To handle the platoon formation tracking control on curved roads,  we decouple the control design strategy into path following and formation control problems. Specifically, on one side, lateral control laws are designed for each vehicle ensuring a geometrical convergence towards the reference path regardless of the translational velocity; on the other side, longitudinal control laws are proposed for each vehicle to achieve the desired relative arc length and velocity with respect to its neighboring vehicle.

One of the key distinctions of the proposed work is that we adopt a novel concept of constructive barrier feedback, first presented in \cite{Zhiqi2023Constructive}, for reactive collision avoidance. The constructive barrier feedback exploits \textit{divergent flow} \cite{bhagavatula2011optic}, a natural feature inspired by insects and birds, to prevent collisions while effectively achieving the primary control objective. 
In this work as well as in our preliminary work \cite{XiaoPlatoonform}, constructive barrier feedback is exploited as an additive term to the nominal lateral and longitudinal controllers, which effectively avoids collision between neighboring vehicles and the road edges without compromising the nominal control objectives. Building on our preliminary results in \cite{XiaoPlatoonform}, which considered simplified vehicle dynamics (second-order systems) and straight roads, this work extends the previous findings by addressing a group of CAVs modeled as nonholonomic second-order kinematic bicycle systems operating on curved roads. The incorporation of constructive barrier feedback enables the design to take an explicit state-feedback form, offering a simple and elegant structure with low computational complexity, while providing formal guarantees for safety invariance and stability.

Under the proposed control methods, all vehicles accurately track the desired reference path and converge to the desired formation while maintaining safe distances from both their platoon predecessor and road boundaries, provided the initial conditions are safe. The effectiveness of the proposed approach is demonstrated through theoretical analysis, simulation studies, and experimental validation using connected miniature vehicles.

The rest of the paper is structured as follows. Section \ref{sec:Pre} provides preliminary results about Constructive barrier feedback. In Section \ref{sec.problem}, we define the platoon formation problem and present the vehicle model used for control design in this study. Section \ref{sec.control} outlines the formulation of the proposed method. The stability property of the method is further stated in Section \ref{sec.stability}. Section \ref{sec.result} presents simulation results aimed at validating the effectiveness of the proposed method. Additionally, we compare these results with a baseline, represented by the nominal control without the constructive barrier feedback component. Experimental studies based on miniature vehicles are conducted and presented in Section \ref{sec.experiment}. Finally, concluding remarks and insights for future research are summarized in Section \ref{sec.conclusion}. 

\section{Preliminary} \label{sec:Pre}
\subsection{Constructive barrier feedback for collision avoidance}
In this section, we will recall the concept of constructive barrier feedback proposed in \cite{Zhiqi2023Constructive} for collision avoidance of a leader-follower structure in which each agent dynamics is described as a double integrator as follows
	\begin{equation}\label{eq:double integrator}
		\left\{
		\begin{aligned}
			\dot{\mathbf{p}}_i&=\mathbf{v}_i\\
			\dot{\mathbf{v}}_i&=\mathbf{u}_i
		\end{aligned}
		\right.
	\end{equation}
where  $ \mathbf{p}_i\in\mathbb{R}^m (m\ge 2)$ is the position,  $ \mathbf{v}_i\in\mathbb{R}^m$ is the velocity of each agent $i$, respectively, and $\mathbf{u}_i\in\mathbb{R}^m$ is the input acceleration. 

 

To get an effective reactive collision avoidance without affecting the stability property of the nominal controller, feedback controller $u_i$ is designed in \cite{Zhiqi2023Constructive} as:
	\begin{equation}\label{eq:ui}
		\mathbf{u}_i=\mathbf{u}_i^n+\mathbf{u}_i^c, 
	\end{equation}
	where $\mathbf{u}_i^n$ is the nominal control input ensuring the asymptotic (or the exponential) convergence of the states $(\mathbf{p}_i,\mathbf{v}_i)$ to the desired trajectory. $\mathbf{u}_i^c$, is a dissipative \textit{control barrier feedback}  slowing down the relative velocity of agent $i$ in the direction of the neighbor agent $j$, $\mathbf{g}_{ij}:=\frac{\mathbf{p}_i-\mathbf{p}_j}{\|\mathbf{p}_i-\mathbf{p}_j\|}$, without compromising the stability nature of the nominal control action
 \begin{equation}\label{eq:cbf}
		\mathbf{u}_i^c=\mathbf{g}_{ij}\frac{\dot d_{ij}}{d_{ij}}
	\end{equation}
 where $d_{ij}:=\|\mathbf{p}_i-\mathbf{p}_j\|-r$ and the divergent flow
 $\frac{\dot d_{ij}}{d_{ij}}$
can be obtained directly from the optical flow using visual information \cite{rosa2014opticalflow}, or estimated from the measure of $d_{ij}$ \cite{hua2010telemetric_measurements}.

To illustrate the obstacle avoidance principle employed in this context, let's consider a 2-agent system. Using the above definitions of $d=d_{ij}$ and $\dot{d}=\dot d_{ij}=\mathbf{g}_{ij}^\top (\mathbf v_i-\mathbf v_j)$, it is straightforward to verify that:
	\begin{equation} \label{ddot_d}
		\ddot d=-k_o\frac{\dot d }{d} -\alpha_i(t) 
	\end{equation}
	with $\alpha_i(t)=-\frac{\|\mathbf{\pi}_{\mathbf{g}_{ij}}(\mathbf{v}_i-\mathbf{v}_j)\|^2}{d+r} -\mathbf{g}_{ij} ^\top(\mathbf{u}_i-\mathbf{u}_{j})$.
	The barrier effect of $\mathbf{u}_i^c$, is announced  in the following technical lemma:
	\begin{lem}\label{lem:boundness of OF}
			Given the dynamics \eqref{ddot_d}
			with $k_o$ a positive gain and $\alpha_i(t)$ a continuous and bounded function. Then for any initial condition satisfying $d(0)>0$ and $\phi(0)=\frac{\dot d(0)} {d(0)}$ bounded, the following assertions hold:
			\begin{enumerate}
				\item $d$ remains positive, $\forall t\ge 0$.
				\item $d$ converges to zero as $t\to \infty$ if and only if $\lim_{t \to \infty} \int^\top_0 \alpha(\tau) d\tau \to +\infty$. 
				\item If $d$ converges to zero, then $\dot{d}$ is bounded and converges to zero, and $\phi(t)$ remains bounded, $\forall t\ge 0$. Furthermore, if $\alpha_i(t)$ converges to a positive constant $\alpha^0>\epsilon >0$, then $\frac{\dot d}{d}\to -\frac{\alpha^0}{k_o}$ and hence $\ddot d$ converges to zero.
			\end{enumerate}
	\end{lem}

Proof of the lemma is given in \cite{Zhiqi2023Constructive}. This lemma shows the safety invariance property, such that, as long as the initial distance $d(0)$ is positive, $d$ will never cross zero for all times as long as the nominal controller $\mathbf{u}^n_i$, the neighboring agent input $\mathbf{u}_{j}$, and the relative velocity $\mathbf{v}_i-\mathbf{v}_j$ are continuous and bounded.
\section{problem formulation}\label{sec.problem}
\begin{figure}[t]
	\centering	\centerline{\includegraphics[trim={10cm 7cm 10cm 9cm},clip,width=1\linewidth]{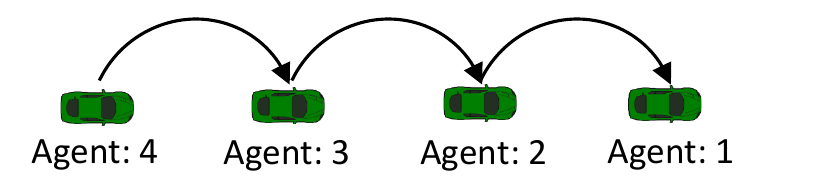}}
	\caption{Interaction topology for a 4-agent platoon formation scenario. The arrow indicates the information access for each agent $i$ from its neighboring agent $i-1$.}
	\label{fig:topology}
\end{figure}

This paper considers a platoon formation tracking control problem for $n$ vehicles on a curved multi-lane road. The concept of constructive barrier feedback is exploited in the control design to ensure collision avoidance between neighboring vehicles and the road edges during the platoon formation process. Each vehicle in the platoon is assumed to be connected to its neighboring agents under a directed graph topology as described in the following assumption:
\begin{assumption}\label{ass:topology}
		The topology $\mathcal G$ is fixed and described by an acyclic digraph with a single directed spanning tree, as shown in Fig. \ref{fig:topology}.
		Without loss of generality, agents are numbered (or can be renumbered) such that agent $1$ is the leader, i.e.,  $\mathcal{N}_1= \varnothing$,  all other agents $i, \ i\ge 2$ are followers whose neighboring set is $\mathcal{N}_i = \{i-1\}$.
\end{assumption}

For vehicle dynamics, instead of a simple second-order integrator model considered as in the previous work \cite{Zhiqi2023Constructive}, the motion of each vehicle $i$ is described by a nonholonomic second-order kinematic bicycle model
\begin{equation}\label{eq.kinematic}
    \left\{
    \begin{aligned}
        &\dot{p}_{xi} = v_i\cos\theta_i\\
        &\dot{p}_{yi} = v_i\sin\theta_i\\
        &\dot{v}_i = a_i\\
        &\dot{\theta}_i = v_i\chi_i
    \end{aligned}
    \right.
\end{equation}
where $\mathbf{p}_i = [p_{xi}\; p_{yi}]^\top \in \mathbb R^2$ is the center position of the rear axle for vehicle $i$ expressed in a common fixed world frame $\mathcal I$. $\theta_i$ and $v_i$ indicate the orientation and speed of vehicle $i$, respectively. The control input are acceleration $a_i$ and input curvature $\chi_i=\frac{\tan\delta_i}{L_i}$, where $\delta_i$ and $L_i$ denote steering angle and wheel base of vehicle $i$. 

The platoon formation tracking control on a curved road can be considered a combination of path-following and formation-control problems. The control design strategy, hence, can be decoupled as i) designing lateral control laws for follower vehicles, ensuring a geometrical convergence towards the reference path regardless of the translational velocity $v_i$; ii) deriving longitudinal control laws for the follower vehicles to achieve desired relative arc length with its neighboring vehicle 
on the reference path while maintaining the same desired speed as its neighboring vehicle $i-1$. 

\begin{figure}[t]
	\centering	\centerline{\includegraphics[trim={7cm 3cm 5.5cm 4cm},clip,width=1\linewidth]{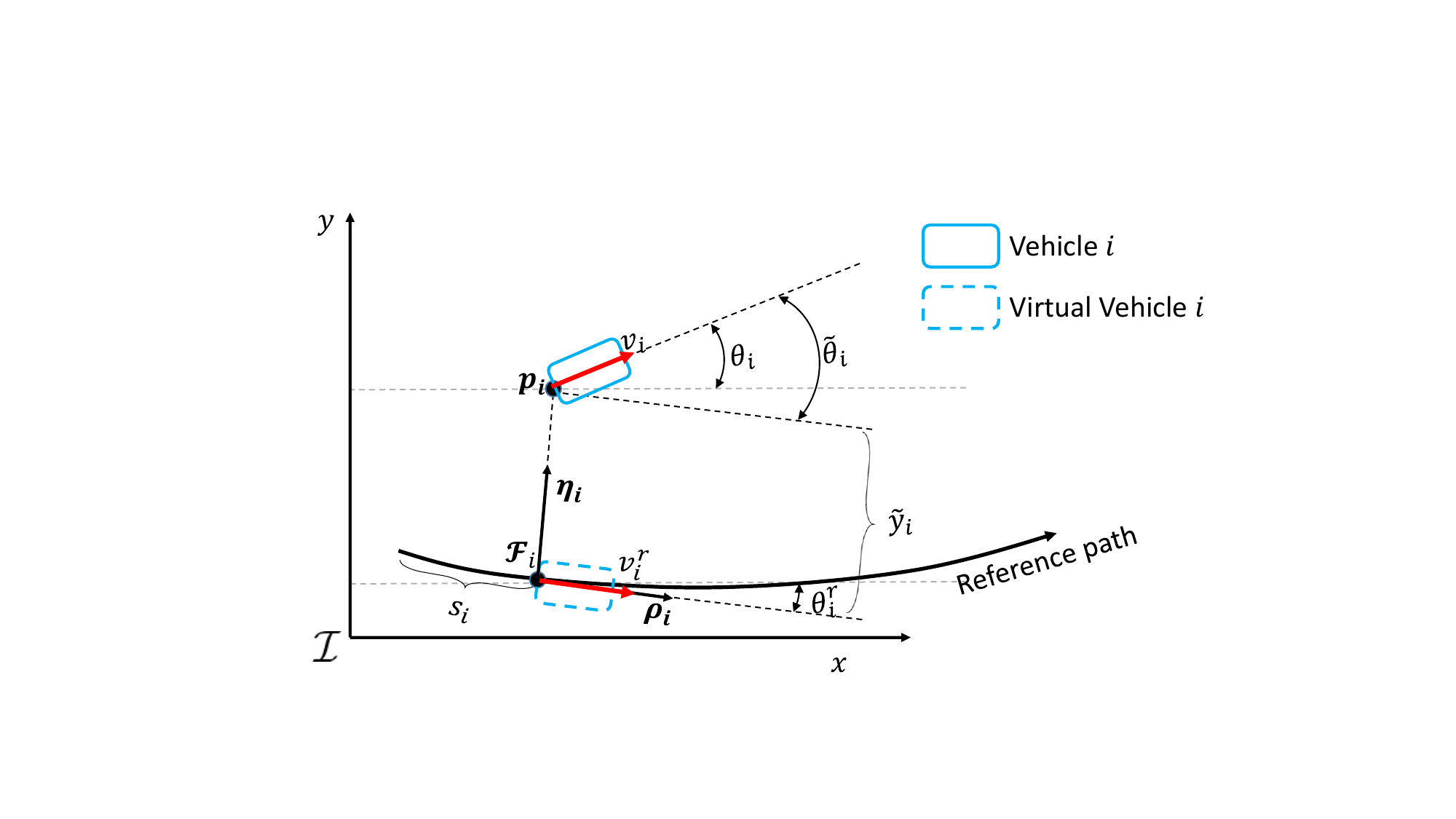}}
	\caption{Kinematic bicycle model for 2-dimensional vehicle motion in both global and Frenet frame. The rectangle of the solid blue line indicates the actual vehicle $i$ and the rectangle of the dashed blue line represents the corresponding virtual vehicle on the path. }
	\label{fig.frame}
\end{figure}

To conveniently handle the vehicle control problem along the given reference path, we transform the expression of the vehicle dynamics from the fixed world frame  $\mathcal I$ to a reference frame aligned with the path $\mathcal F_i$ similarly to the path following control problems in \cite{Frenet}. The path-aligned frame is analogous to the Frenet frame with the distinction that the normal direction of the path is not necessarily oriented toward the path's curve. It can be seen as a kinematically equivalent fictitious vehicle's virtual frame on the path with longitudinal and lateral axes $\boldsymbol{\rho}_i$ and $\boldsymbol \eta_i$, respectively, as shown in Fig. \ref{fig.frame}. Since the virtual vehicle $i$ is defined by the orthogonal projection of the actual vehicle $i$'s position onto the reference path, using the result from \cite{Frenet}, the velocity of the virtual vehicle satisfies the following condition
\begin{equation}\label{eq.vr}
v_i^r=\frac{v_i\cos\Tilde{\theta}_i}{1-\chi_i^r\Tilde{y}_i}\\
\end{equation}
where $\tilde{y}_i$ and $\tilde{\theta}_i = \theta_i - \theta^r_i$ denote the lateral displacement error and orientation error of vehicle $i$ with respect to the virtual frame $\mathcal F_i$ with following dynamics

\begin{equation}\label{eq.dyna_frenet}
    \left\{
    \begin{aligned}
        &\dot{\Tilde{y}}_i = v_i\sin\Tilde{\theta}_i\\
        &\dot{\Tilde{\theta}}_i = v_i\left(\chi_i - \frac{\chi_i^r\cos\Tilde{\theta}_i}{1-\chi_i^r\Tilde{y}_i}\right).
    \end{aligned}
    \right.  
\end{equation}
where $\chi_i^r$ is curvature of the virtual vehicle $i$ on reference path. The curvature $\chi_i$ is the control input to be designed as the lateral controller to drive $(\tilde y_i,\tilde \theta_i)$ to zero. Note that equation \eqref{eq.vr} and \eqref{eq.dyna_frenet} are well defined as long as the projection onto the path is unique, that is  
$|\tilde y_i|<\frac 1 {\chi_i^r}$.

Note that in the classical path-following problem, translational and orientation control are typically decoupled to stabilize the equilibrium $(\tilde{y},\tilde{\theta})=(0,0)$ independently from the translational motion. In the proposed approach, longitudinal control is an additional objective aimed at tracking the preceding vehicle. In this context and in contrast to the path-following literature, we reverse the constraint in \eqref{eq.vr} by first imposing the desired dynamics of the virtual vehicle $v_i^r$ and then use this constraint to determine the control action for the actual vehicle:
\begin{equation}\label{eq.vi}
v_i=v_i^r\frac{1-\chi_i^r\Tilde{y}_i}{\cos\Tilde{\theta}_i}\\
\end{equation}
The reversed constraint is well-defined given $|\Tilde{\theta}_i|<\frac{\pi}{2}$. Hence, the design strategy focuses first on the virtual acceleration control input for the virtual follower vehicle $i$. Analyze the dynamics of arc length and velocity $(s_i,v_i^r)$ of the virtual vehicle $i$, 

\begin{equation}\label{eq.dyna_frenet_lon_ref}
    \left\{
    \begin{aligned}
        &\dot{s}_i = v_i^r\\
        &\dot{v}^r_i = a^r_i
    \end{aligned}
    \right.  
\end{equation}
where $a_i^r$ is the virtual acceleration control to be designed to drive $(s_i,v_i^r)$ to the desired arc length and velocity $(s_i^*,v_i^*)$. 
Note that as long as  the constraints $|\tilde y_i|<\frac 1 {\chi_i^r}$ and $|\tilde{\theta}_i|<\frac \pi 2$ associated with equations \eqref{eq.vr} and \eqref{eq.vi} are fullfilled respectively, one can recover the actual input $a_i$ from $a_i^r$, $\chi_i$, the state variables $(v_i,\tilde \theta_i, \tilde y_i)$ and the reference path information $(\chi_i^r,\dot \chi_i^r)$ using the derivative of \eqref{eq.vi}:

\begin{equation}\label{eq.ar2a1}
\begin{aligned}
    a_i = & \frac{1}{\cos\Tilde{\theta}_i}\biggl(
    a_i^r(1-\chi_i^r\Tilde{y}_i)+v_i\sin\Tilde{\theta}_i \dot{\Tilde{\theta}}_i\\
    &-v_i^r\left(\dot \chi_i^r \Tilde{y}_i+\chi_i^r\dot{\Tilde{y}}_i\right)\biggr)
\end{aligned}
\end{equation}
Now, since for any continuous function $f(t)$ and $g(t)$ having $\dot{f}(t)=\dot{g}(t)$ does not imply that $f(t)=g(t)$, one suggests the following modification to satisfy the constraint \eqref{eq.vi} :
\begin{equation}\label{eq.ar2a}
\begin{aligned}
    a_i = & \frac{1}{\cos\Tilde{\theta}_i}\biggl(
    a_i^r(1-\chi_i^r\Tilde{y}_i)+v_i\sin\Tilde{\theta}_i \dot{\Tilde{\theta}}_i\\
    &-v_i^r\left(\dot \chi_i^r \Tilde{y}_i+\chi_i^r\dot{\Tilde{y}}_i\right)\biggr)-k\left (v_i^r\frac{1-\chi_i^r\Tilde{y}_i}{\cos\Tilde{\theta}_i}-v_i \right)
\end{aligned}
\end{equation}
with $k$ a positive real number.

\begin{figure}[t]
	\centering	\centerline{\includegraphics[trim={7cm 3cm 5cm 4cm},clip,width=1\linewidth]{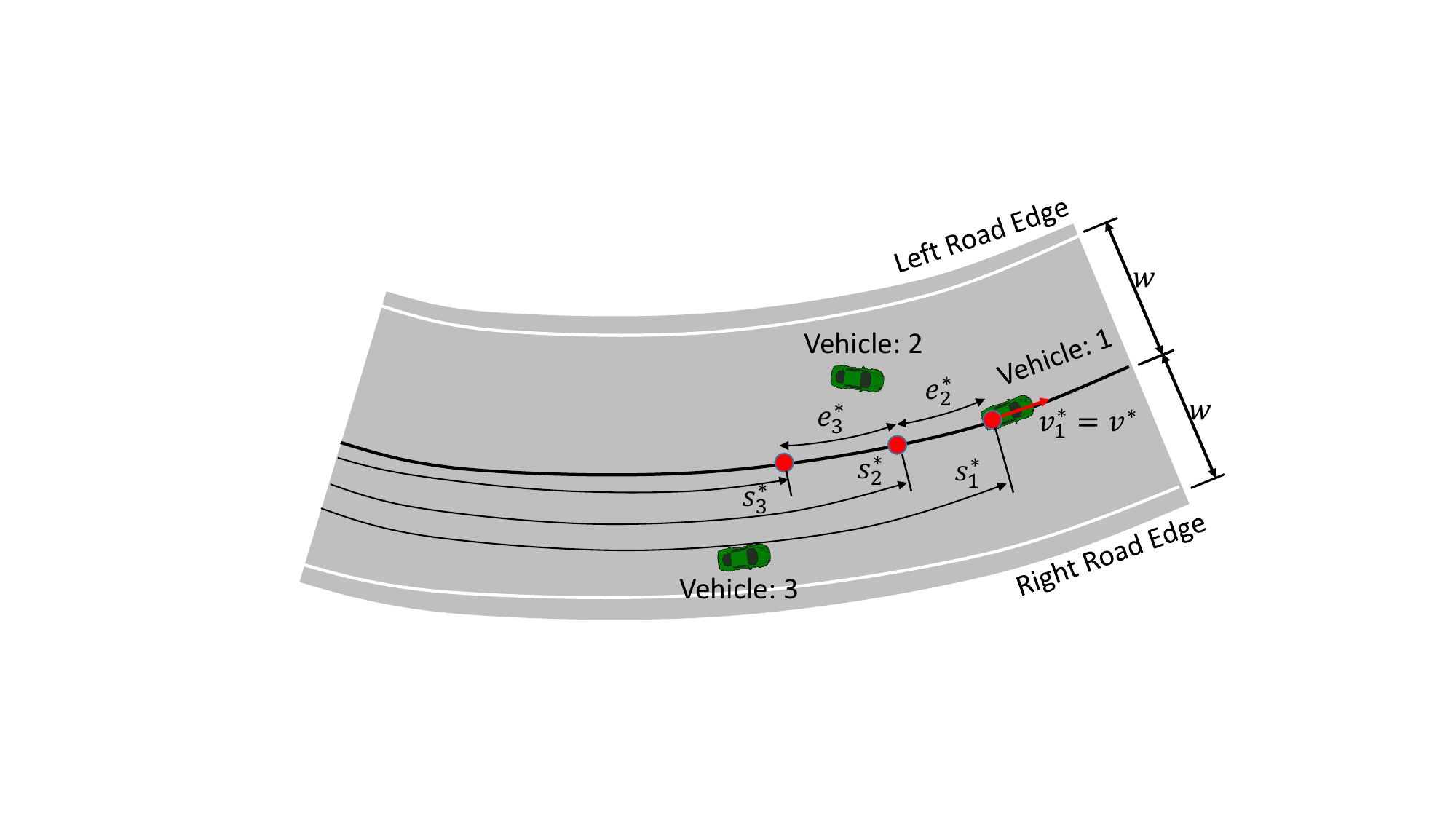}}
	\caption{Road layout to illustrate the desired platoon formation.}
	\label{fig.formation}
\end{figure}

With the above description, the desired platoon formation is defined as follows (see Fig. \ref{fig.formation}):
\begin{assumption}\label{ass.platoon_form}
    The leader vehicle, agent 1, is independently controlled such that it is already in its desired configuration: $\tilde y_1=0$, $\tilde{\theta}_1=0$, $v_1=v_1^*=v^*>0$, and $s_1=s_1^*$. The desired configuration of the follower vehicle $i$, $i\geq2$ is on the reference path (i.e., $\tilde y_i=0$ and $\tilde{\theta}_i=0)$ while following its platoon predecessor $i-1$ with desired velocity $v_i^*=v_{i-1}^*>0$ and desired relative arc length $e_i^* := s^*_{i-1} - s^*_{i}>L_i+\epsilon>0$, where $\epsilon>0$ is a predefined safe margin.
\end{assumption}

Besides achieving the desired platoon formation, the group of vehicles has to follow the traffic rules and, hence, should not cross the road boundaries. The assumption below is to describe the road environment and its relationship with the reference path. 
\begin{assumption}\label{ass.road_shape}
     The reference path is parameterized by a smooth arc length $s$ such that the curvature of the path $\chi^r(s)$ is continuous and differentiable with respect to $s$. Without loss of generality, the platoon's reference path coincides with the road's center line. As shown in Fig. \ref{fig.safety}, The left and right road edges are parallel to the road center line with a constant lateral width offset distance $w$ such that $\frac{1}{\chi^r}>w>\epsilon_w>0$ for all $s$, where $\epsilon_w>0$ is a predefined lateral safe margin. In addition, for any two distinct points $\mathbf{p}_i$ and $\mathbf{p}_j$ projecting to the reference path with arc length $s_i$ and $s_j$,  $|s_i-s_j|>  \epsilon>0$ implies $||\mathbf{p}_i - \mathbf{p}_j||_2 >  \epsilon>0$
\end{assumption}

The following last assumption assigns the order of the vehicles' index according to their initial arc length projecting on the reference path. This condition will be used later for collision avoidance 
 control design between neighboring vehicles.
\begin{assumption}\label{ass.vehicle_index}
    Provided $s_i(0)-s_j(0) \ne 0, \forall i,j\in \mathcal{V} = {1,2,...,n}$, the vertex set of $n$-vehicle system is assigned such that $s_{i-1}(0)-s_i(0)>0,\forall i\geq 2$.
\end{assumption}

Given the above ingredients, the safe platoon formation tracking control problem is formally formulated as follows: 
\begin{prob}\label{prob.1}
    Under Assumptions \ref{ass:topology} - \ref{ass.vehicle_index}, design lateral and longitudinal controller $\chi_i$ and $a_i^r$, respectively for follower vehicles $i\ge 2$, under which the multi-vehicle system achieves the desired platoon formation described in Assumption \ref{ass.platoon_form} while guaranteeing collision-free to road edges and between neighboring vehicles for all the time 
\end{prob}

\section{Safe platooning and merging controller design}\label{sec.control}
\subsection{Lateral controller design}
From the previous problem formulation in Section \ref{sec.problem}, the lateral control design aims to ensure the follower vehicle a geometrical convergence towards the reference path while guaranteeing collision-free with the road edges. 
Inspired by the work \cite{Zhiqi2023Constructive}, the safe lateral controller for each follower vehicle $i\ge 2$ is proposed as two parts:
\begin{equation}\label{eq.control_lat}
\begin{aligned}
    \chi_i =& \chi_i^n+\chi_i^c
\end{aligned}
\end{equation}
where $\chi_i^n$ is the nominal controller that ensures the stabilization of the origin of error variable $(\tilde y_i,\tilde \theta_i)$ and $\chi_i^c$ is the constructive barrier feedback to be designed for avoiding collision with the road edges.

Hence, the nominal lateral controller for the follower vehicle $i\geq2$ is designed as 
\begin{equation}\label{eq.control_lat_n}
\begin{aligned}
    \chi_i^n = -k_1\frac{\sin\Tilde{\theta}_i}{\Tilde{\theta}_i}\Tilde{y}_i - k_2\sign(v_i)\Tilde{\theta}_i
    +  \frac{\chi_i^r\cos\Tilde{\theta}_i}{1-\chi_i^r\Tilde{y}_i}    
\end{aligned}
\end{equation}
where $k_1$ and $k_2$ are positive gains, the first two terms are designed to correct lateral error $\tilde{y}_i$ and orientation error $\tilde{\theta}_i$ respectively, and the last is the feedforward term used to track the reference curvature $\chi_i^r$. 

To prevent vehicles from crossing the road edges, we define the lateral safety distance to the left and right road edges, respectively, as
\begin{equation}\label{eq.deta}
    d^{\eta_L}_i = w-\Tilde{y}_i-\epsilon_w, \ d^{\eta_R}_i = w+\Tilde{y}_i-\epsilon_w
\end{equation}
where $w$ is the distance to the road edge from the reference path and $\epsilon_w>0$ is a predefined safe margin as seen in Fig. \ref{fig.safety}. 

It is straightforward to verify that $ d^{\eta_L}_i> 0$ and  $ d^{\eta_R}_i> 0$ implies that i) vehicle $i$ stays within the left and right road edges with a predefined safety margin $\epsilon_w>0$, ii)$|\Tilde{y}_i|< w-\epsilon_w < \frac{1}{|\chi_i^r|}$, based on Assumption \ref{ass.road_shape}, which ensures that $v_i^r$ in \eqref{eq.vr} is well defined.

Analogous to the divergent flow used in \eqref{eq:cbf}, the lateral constructive barrier feedback $\chi_i^c$ is designed as: 
\begin{equation}
\begin{aligned}
    \chi_i^c =- k_3 \left (\frac{{d^{\eta_R}_i}^{'}}{d^{\eta_R}_i}-\frac{{d^{\eta_L}_i}^{'}}{d^{\eta_L}_i}\right)
\end{aligned}
\end{equation}
where the notation $(.)^{'}$ denotes derivative with respect to the curvilinear abscissa $\sigma =\int_0^t |v_i(\tau)| d\tau$, i.e., ${d_i^{\eta_L}}^{'} =\frac{d}{d\sigma} d^{\eta_L}_i=-\sign(v_i)\sin\Tilde{\theta}_i$ and ${d_i^{\eta_R}}^{'} =\sign(v_i)\sin\Tilde{\theta}_i$. Hence $\chi_i^c$ can be represented as
\begin{equation}\label{eq.control_lat_o}
\begin{aligned}
    \chi_i^c =-k_3\left(\frac 1 {d^{\eta_L}_i}+\frac 1 {d^{\eta_R}_i}\right)\sign(v_i)\sin\Tilde{\theta}_i
\end{aligned}
\end{equation}
Instead of using the time derivative of the distance as in \cite{Zhiqi2023Constructive}, the innovation here is to use the derivative of the distance with respect to the new variable $\sigma$ to avoid longitudinal translational velocity $v_i$ appearing in the divergent flow, decoupling the lateral and longitudinal control design.

\



\subsection{Longitudinal controller design}
\begin{figure}[t]
	\centering	\centerline{\includegraphics[trim={6cm 4cm 4.5cm 3cm},clip,width=1\linewidth]{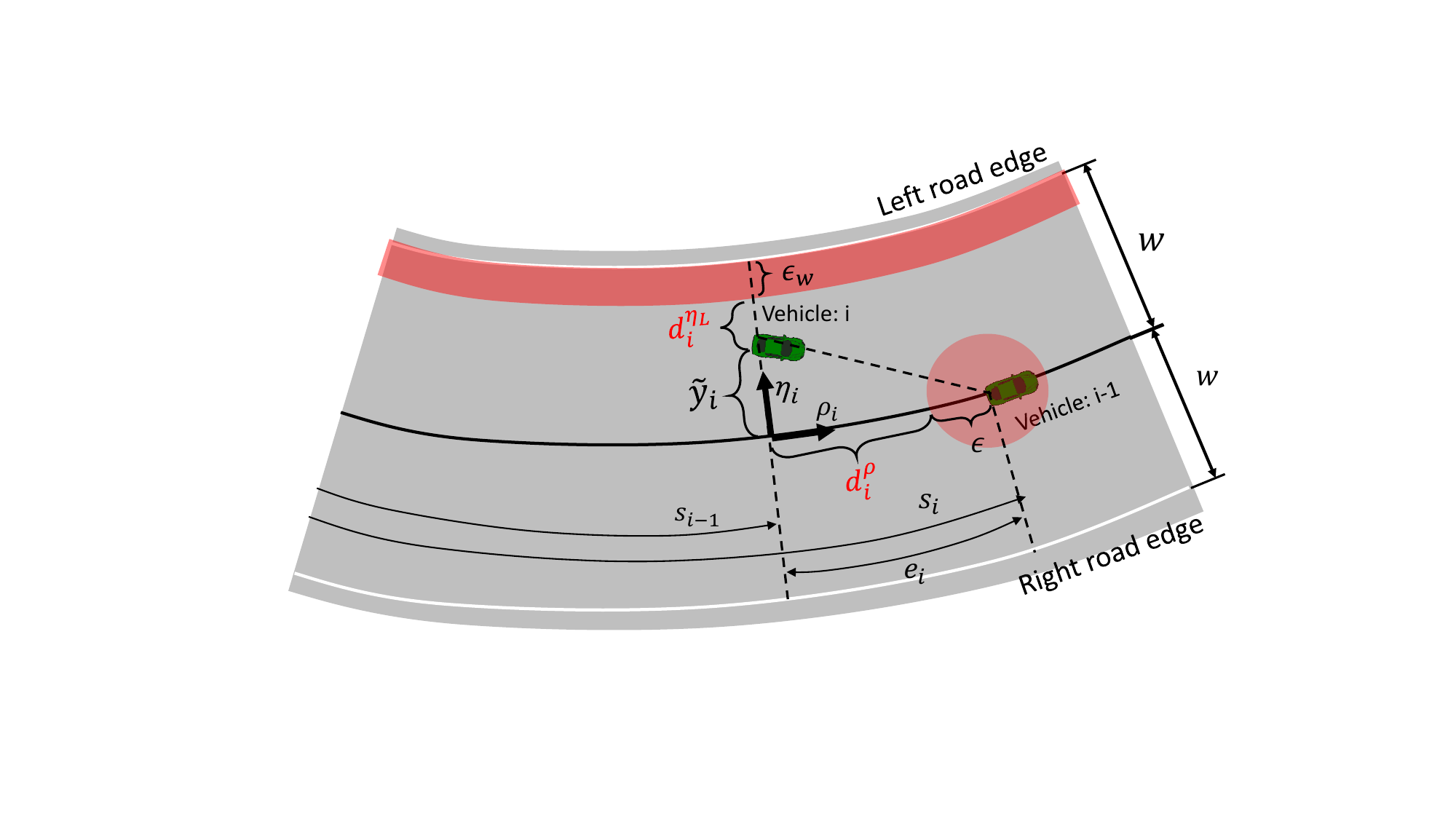}}
 	\caption{The lateral safety distance $d_i^{\eta_L}$ of vehicle $i$ with respect to road edges and the longitudinal safety distance $d_i^\rho$ of vehicle $i$ with respect to its predecessor $i-1$. }
	\label{fig.safety}
\end{figure}

For the longitudinal control, the design is initiated over the virtual vehicle. The virtual acceleration control input $a_i^r$ in \eqref{eq.dyna_frenet_lon_ref} is designed as two parts similar to the lateral controller \eqref{eq.control_lat}
\begin{equation}\label{eq.lon_control_vir}
    a_i^r = a_i^{n}+a_i^{c}.
\end{equation}
The nominal controller $a_i^{n}$ is responsible for driving the virtual vehicle’s longitudinal state $(s_i,v_i^r)$ toward the desired platoon configuration $(s_i^*,v_i^*)$. It is defined as
\begin{equation}\label{eq.lon_control_vir_n}
    a_i^{n} = k_4\Tilde{e}_i+k_5\nu_i+a_{i-1}^r
\end{equation}
where $\Tilde{e}_i: = e_i - e_i^*$ represents the relative longitudinal position error, with $e_i = s_{i-1}-s_{i}$ denoting the difference in arc length between virtual agent $i$ and $i-1$. The desired inter-vehicle spacing is given by $e_i^*$. The term $\nu_i := (v^r_{i-1} - v^r_i)$ is the relative velocity error, and $k_4$ and $k_5$ are positive gains. 

To prevent collision between vehicle $i$ and $i-1$, the longitudinal safety distance $d_i^{\rho}$ between virtual vehicles $i$ and $i-1$ is introduced as 
\begin{equation}
    d_i^{\rho}: = e_i -\epsilon
\end{equation}
where $\epsilon>0$ is the safety margin. See Fig. \ref{fig.safety} for a graphic view of the safety distance. Recall Assumption \ref{ass.road_shape}, one concludes that collision-free between actual vehicles $i$ and $i-1$ is guaranteed, i.e., $||\mathbf{p}_{i-1}-\mathbf{p}_{i}||_2  >\epsilon> 0$, as long as the collision-free between virtual vehicles holds, i.e., $ d_i^{\rho}>0$.

Finally, the longitudinal constructive barrier feedback for collision avoidance between neighboring virtual vehicles is designed as 
\begin{equation}\label{eq.lon_control_vir_o}
    a_i^{c} = k_6\frac{\dot d_i^{\rho}}{ d_i^{\rho}}
\end{equation}
where $\dot{ d_i^{\rho}} = \nu_i$, and $k_6$ is a positive control gain.

\section{Stability analysis}\label{sec.stability}
Under the proposed control design, the following lemma shows safety guarantees and equilibrium analysis of the longitudinal controller.
\begin{lem}\label{lem.boundedv}
Consider a $n$-vehicle system with the system dynamic \eqref{eq.kinematic} and the virtual longitudinal control \eqref{eq.lon_control_vir}, \eqref{eq.lon_control_vir_n} and \eqref{eq.lon_control_vir_o}. If Assumption \ref{ass:topology} - \ref{ass.vehicle_index} are satisfied, then for any bounded initial conditions such that $d_i^{\rho}(0)>0$ and $a_i^c(0)$ bounded, the following assertions hold $\forall i\in\mathcal{V}/\{1\}$, $\forall t \geq 0$:
    \begin{enumerate}
        \item vehicle $i$ remains collision-free with its neighboring vehicle $i-1$, i.e., $d_i^{\rho}(t)$ remains positive, and the controller $a_i^r$ is well-defined  and  bounded;
        \item the origin of $(\tilde e_i,\nu_i)$ is asymptotically stable;
    \end{enumerate}
\end{lem}
The proof of this lemma is provided in Appendix \ref{app.1}. 

Under both lateral and longitudinal controllers, the multi-vehicle systems' safety invariant properties and stability analysis are provided by the following theorem.
\begin{thm}\label{thmNvehicle}
Consider a $n$-vehicle system with the dynamics \eqref{eq.kinematic} along with the longitudinal controller defined by \eqref {eq.ar2a}, \eqref{eq.lon_control_vir}, \eqref{eq.lon_control_vir_n} and \eqref{eq.lon_control_vir_o} and lateral controller  \eqref{eq.control_lat}, \eqref{eq.control_lat_n} and \eqref{eq.control_lat_o}. If Assumption \ref{ass:topology} - \ref{ass.vehicle_index} are satisfied, then under any bounded initial condition satisfying,  $d_i^{\rho}(0) > 0$, $d^{\eta}_i(0) > 0$, $k_1\tilde y_i^2(0)+\Tilde{\theta}_i^2(0)<(\frac{\pi}{2}-\epsilon_1)^2$, (with $\epsilon_1<<1$ a positive number), $a_i^c(0)$ and $\chi_i^c(0)$ bounded, 
the following assertions hold $\forall i\in\mathcal{V}/\{1\}$, $\forall t \geq 0$:
\begin{enumerate}
    \item $|\tilde \theta_i|<\frac \pi 2$ and $|\tilde y_i|<\frac 1 {\chi_i^r}$, and the $n$-vehicle system remains safe, i.e., $d_i^{\rho}(t)$, $d^{\eta_L}_i(t)$ and $d^{\eta_R}_i(t)$ remain positive;
    \item the control law $\chi_i$ and $a_i$ are well-defined and bounded;
    \item the origins of $(\Tilde{y}_i,\Tilde{\theta}_i) $ and  $(\tilde s_i,\tilde v_i):=(s_i-s_i^*, v_i-v_i^*)$ are asymptotically stable;
    
\end{enumerate}
\end{thm}
The proof of this theorem is provided in Appendix \ref{app.2}.

\section{Simulation results}\label{sec.result}
\begin{figure*}[t]
     \centering
     \begin{subfigure}[b]{0.48\textwidth}
        \centering
            \includegraphics[trim=5.5cm 7.5cm 5.5cm 7.5cm,clip,width=\linewidth]{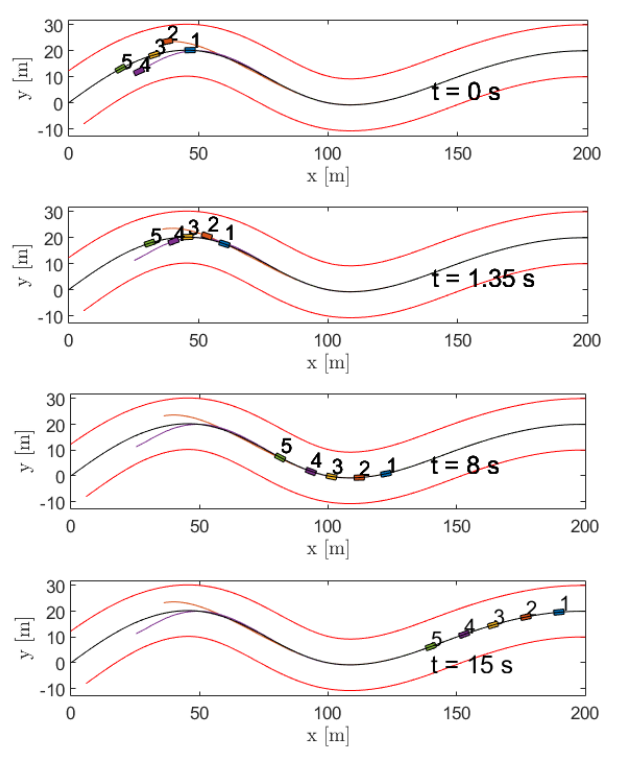}
        \caption{Scenario (A).}
        \label{fig:traj_merg}
     \end{subfigure}
     \begin{subfigure}[b]{0.48\textwidth}
        \centering
            \includegraphics[trim=5.5cm 7.5cm 5.5cm 7.5cm,clip,width=\linewidth]{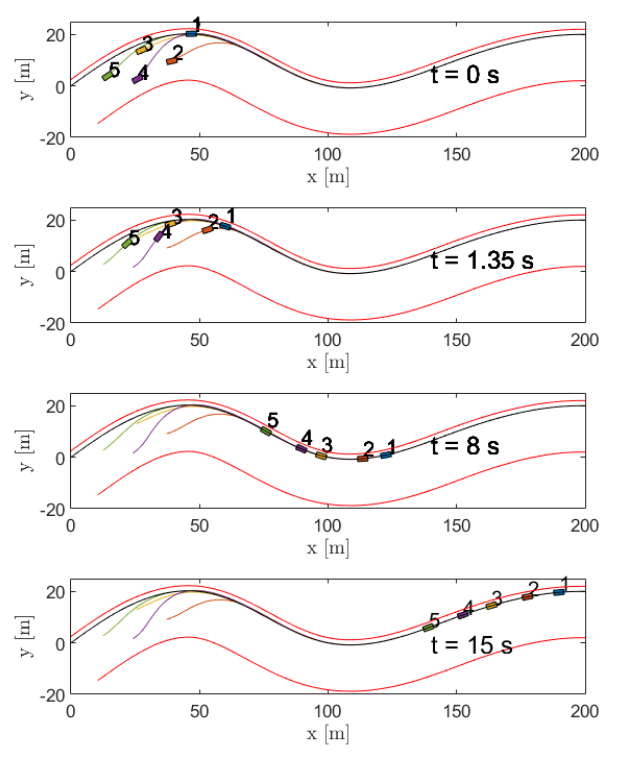}
        \caption{Scenario (B).}
        \label{fig:traj_form}
     \end{subfigure}
    \caption{Time evolution snapshot of the platoon formation process at distinct time points. Two parallel red solid lines are the road edges, the black solid line indicates the desired lane for the platoon, and the color solid lines indicate the vehicle trajectory during the formation process. The left sub-figure shows the result for the merging scenario and the right sub-figure shows the result for the formation scenario.}
        \label{fig:traj_res}
\end{figure*}

To demonstrate the performance of the proposed control method, we design a multilane highway scenario where the shape of the road is generated using the Matlab built-in function $referencePathfrenet()$ with fixed waypoints. In particular, a performance comparison between the proposed safe platoon controller \eqref{eq.control_lat}, \eqref{eq.control_lat_n}, \eqref{eq.control_lat_o}, \eqref{eq.lon_control_vir}, \eqref{eq.lon_control_vir_n}, and \eqref{eq.lon_control_vir_o}. and the nominal platoon controller \eqref{eq.control_lat_n} and \eqref{eq.lon_control_vir_n} (which serves as a baseline control strategy) is provided to emphasize that collision avoidance is ensured without affecting the nominal control objectives. The simulation study is performed for a platoon of five vehicles with two different initial settings, constituting two test scenarios referred to as scenario (A) and scenario (B). The initial conditions are summarized in Table. \ref{tb:merge} and \ref{tb:form} and is visualized in Fig. \ref{fig:traj_res}. 

In both scenarios, the controller gains are set as $k_1 = 0.01$, $k_2 = 0.1$, $k_3 = 0.1$, $k_4 = 0.4$, $k_5 = 0.1$, $k_6 = 2$, and the constant parameters are set as $w = \SI{10}{\metre}$, $\epsilon_w = \SI{1.2}{\metre}$, $\epsilon = \SI{5}{\metre}$, $L_i = \SI{4}{\metre}$, $e^*_i = \SI{14}{\metre}$. $v^*_i = \SI{10}{\metre\per\second}$.

In scenario (A), the reference path is placed on the road centerline as in Fig. \ref{fig:traj_merg}, vehicle states are initiated such that vehicles $1$, $3$, and $5$ are already on the path, while vehicles $2$ and $3$ are placed to merge into the platoon from the left and right side of the path respectively. 

In scenario (B), the reference path is placed parallel to the road centerline with a 2 m offset distance from the left road boundary as in Fig. \ref{fig:traj_form}. Here, vehicles are scattered over the road initially. This setup allows us to evaluate the controller's performance in steering vehicles from different lanes into the desired platoon formation while demonstrating the capability of the controller to keep vehicles on the road and avoid collisions with each other. 

\begin{figure*}[t]
     \centering
     \begin{subfigure}[b]{0.48\textwidth}
        \centering
            \includegraphics[trim=4cm 4cm 4cm 3cm,clip,width=\linewidth]{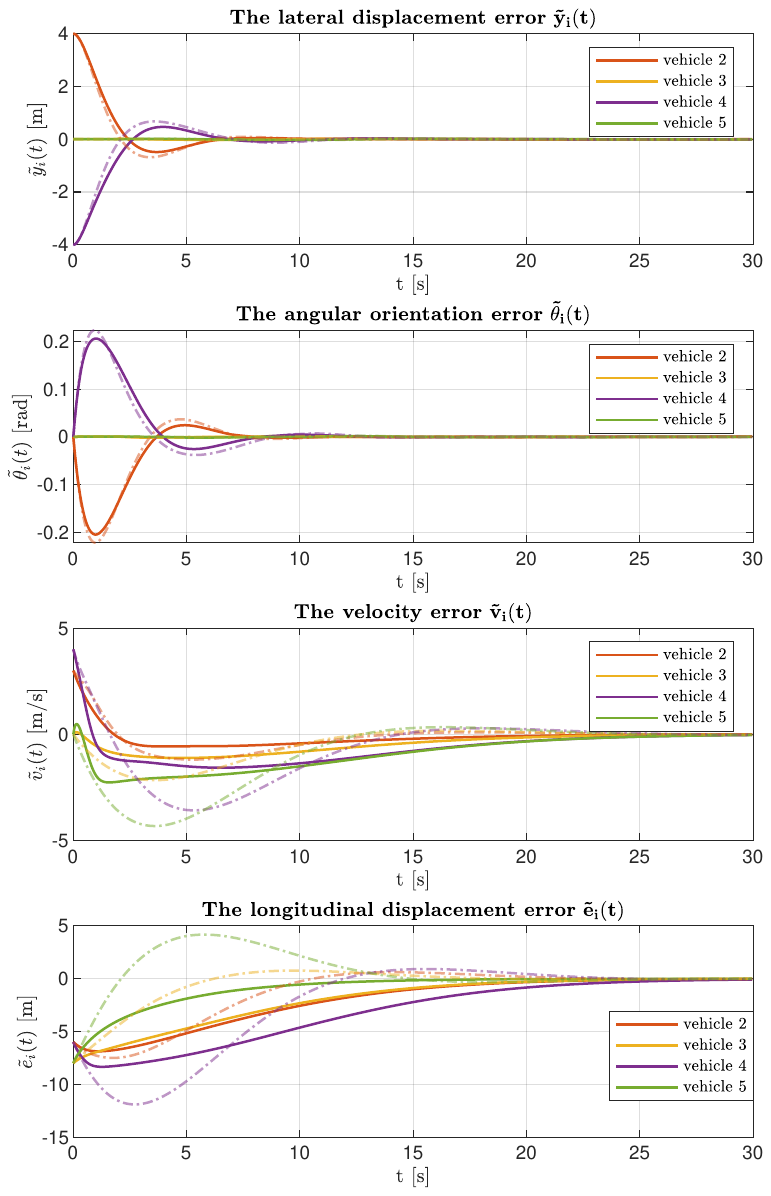}
        \caption{Scenario (A)}
        \label{fig:error_merg}
     \end{subfigure}
     \begin{subfigure}[b]{0.48\textwidth}
        \centering
            \includegraphics[trim=4cm 4cm 4cm 3cm,clip,width=\linewidth]{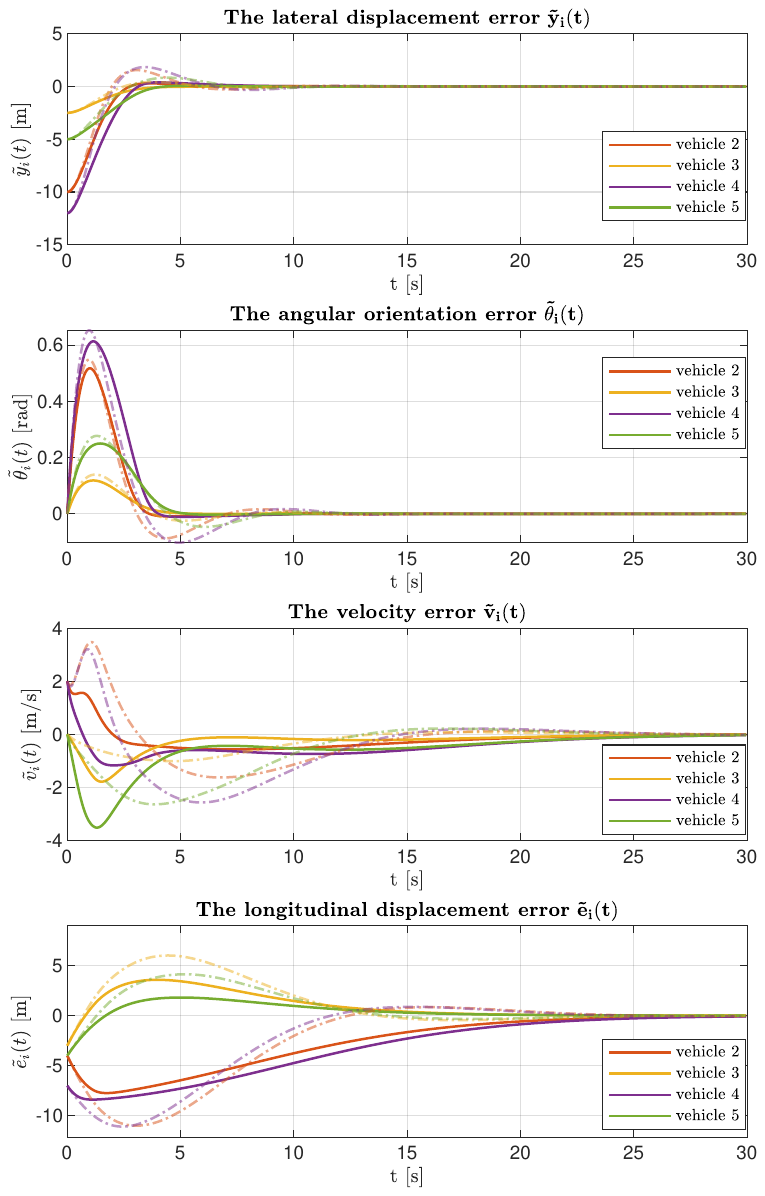}
        \caption{Scenario (B)}
        \label{fig:error_form}
     \end{subfigure}
    \caption{Time evolution of the lateral displacement error $\tilde{y}_i(t)$, the angular orientation error $\tilde{\theta}_i(t)$, the velocity error $\tilde{v}_i(t)$, and the longitudinal displacement error $\tilde{e}_i(t)$ for all $i = \{2,3,4,5\}$ with respective color-coded lines. The solid lines indicate results under the proposed controller \eqref{eq.control_lat}, \eqref{eq.control_lat_n}, \eqref{eq.control_lat_o}, \eqref{eq.lon_control_vir}, \eqref{eq.lon_control_vir_n}, and \eqref{eq.lon_control_vir_o}. The transparent dashed lines indicate the result under the baseline controller \eqref{eq.control_lat_n} and \eqref{eq.lon_control_vir_n}.}
        \label{fig:error_res}
\end{figure*}

The formation evolution under the proposed control method is shown in Fig. \ref{fig:traj_res}, for both scenarios, all vehicles smoothly converge to the desired formation by tracking the reference path and keeping the desired distance with their platoon predecessor. The convergence is more clearly evident in Fig. \ref{fig:error_res}, where the time evolution of the error states $\Tilde{y}_i$, $\Tilde{\theta}_i$, $\Tilde{v}_i$, and $\Tilde{e}_i$ all converge to zero after approximately 20 seconds for both the proposed controller and the baseline. The simulation results indicate that the convergence property of the baseline controller is preserved by the proposed controller. 

\begin{figure*}[t]
     \centering
     \begin{subfigure}[b]{0.48\textwidth}
        \centering
            \includegraphics[trim=4cm 4cm 4cm 3cm,clip,width=\linewidth]{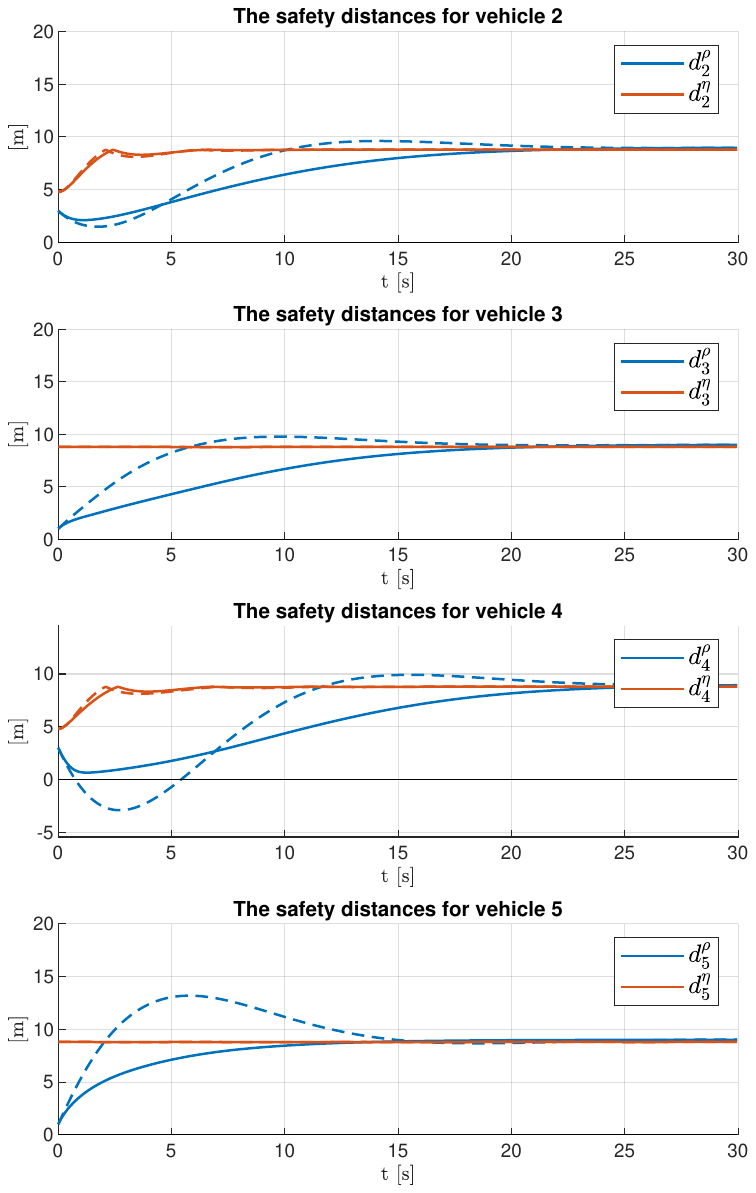}
        \caption{Scenario (A)}
        \label{fig:dist_merg}
     \end{subfigure}
     \begin{subfigure}[b]{0.48\textwidth}
        \centering
            \includegraphics[trim=4cm 4cm 4cm 3cm,clip,width=\linewidth]{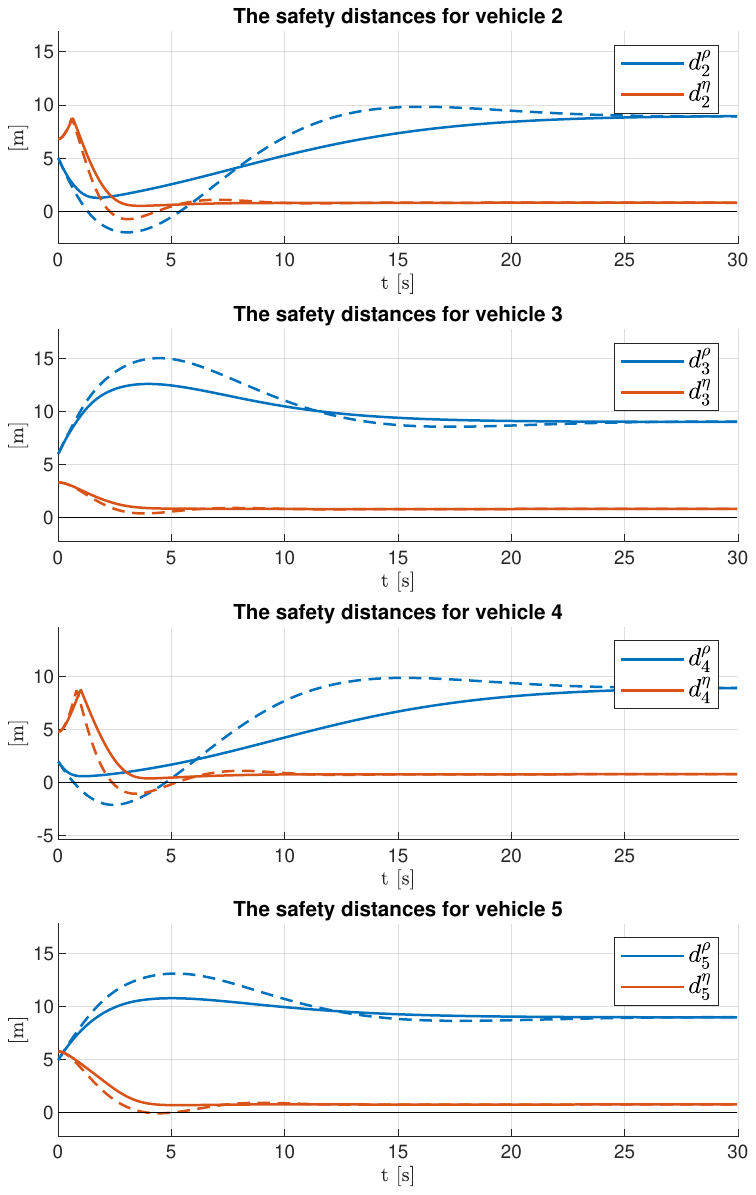}
        \caption{Scenario (B)}
        \label{fig:dist_form}
     \end{subfigure}
    \caption{Time evolution of the safety distance $d_i^\rho(t)$ and $d_{i}^\eta(t)$ for all $i\in\{2,3,4,5\}$. The solid lines indicate the evolution of $d_i^g(t)$ and $d_{i}^\eta(t)$ under the proposed controller \eqref{eq.control_lat}, \eqref{eq.control_lat_n}, \eqref{eq.control_lat_o}, \eqref{eq.lon_control_vir}, \eqref{eq.lon_control_vir_n}, and \eqref{eq.lon_control_vir_o}. The dashed lines indicate the result under the baseline controller \eqref{eq.control_lat_n} and \eqref{eq.lon_control_vir_n}.}
        \label{fig:dist_res}
\end{figure*}

The safety can be measured by $d_i^\rho$ and $d_i^\eta$ as $d_i^\rho\leq 0$ and $d_i^\eta\leq 0 $ indicate vehicle-to-vehicle and vehicle-to-road boundary collision (for simplicity of notation here we denote $d_i^\eta = \min(d_i^{\eta_L}, d_i^{\eta_R})$ for measure of the safety to both sides). As shown in Fig \ref{fig:dist_res} with solid lines, the multi-vehicle system successfully avoids collisions under the proposed controller. In contrast, the multi-vehicle system under the baseline control collides with either neighboring vehicles or the road edges. As shown in Fig. \ref{fig:dist_merg} with dashed lines, for scenario (A), the baseline controller failed to keep $d_4^\rho >0$, where a collision happened between vehicle 4 and vehicle 3. For scenario (B) as shown in Fig. \ref{fig:dist_form}, the baseline failed to keep $d_2^\rho>0$, $d_2^\eta>0$, $d_4^\rho>0$, $d_4^\eta>0$, and $d_5^\eta>0$, indicating that vehicles 2, 4, and 5 crossed the road boundary. In addition, vehicle 2 collided with vehicle 1 and vehicle 4 collided with vehicle 3 in scenario (B).

An animation of the simulation results can be found at \url{https://bit.ly/platoon_formation_curved_road}., where the performance between the proposed controller and the baseline is compared and visualized. In summary, from the simulation comparisons, one concludes that the proposed safe controller efficiently achieves the nominal formation tracking objective while avoiding collision between neighboring vehicles and road edges with reasonable control inputs.

\begin{table}[t]
\renewcommand{\arraystretch}{1.3}
\begin{center}
\caption{Initial vehicle states for scenario (A)}\label{tb:merge}
\begin{tabular}{cccccc}
\hline \hline
Vehicle & $s_{i}$ & $\Tilde{y}_{i}$ & $\Tilde{\theta}_i$ & $v_{i}$ \\
 & [$\si{\metre}$] & [$\si{\metre}$] & [$\si{\radian}$] & [$\si{\metre\per\second}$]\\
\hline
1 & 50 & 0 & 0 & 10\\
\hline
2 & 42 & 4 & 0 & 13\\
\hline
3 & 36 & 0 & 0 & 10\\
\hline
4 & 28 & -4 & 0 & 16\\
\hline
5 & 22 & 0 & 0 & 10\\
\hline
\end{tabular}
\end{center}
\end{table}

\begin{table}[t]
\renewcommand{\arraystretch}{1.3}
\begin{center}
\caption{Initial vehicle states for scenario (B)}\label{tb:form}
\begin{tabular}{cccccc}
\hline \hline
Vehicle & $s_{i}$ & $\Tilde{y}_{i}$ & $\Tilde{\theta}_i$ & $v_{i}$ \\
& [$\si{\metre}$] & [$\si{\metre}$] & [$\si{\radian}$] & [$\si{\metre\per\second}$]\\
\hline
1 & 50 & 0 & 0 & 10\\
\hline
2 & 40 & -10 & 0 & 12\\
\hline
3 & 29 & -2.5 & 0 & 10\\
\hline
4 & 22 & -12 & 0 & 12\\
\hline
5 & 12 & -5 & 0 & 10\\
\hline
\end{tabular}
\end{center}
\end{table}

\begin{figure*}[t]
     \centering
     \begin{subfigure}[b]{0.48\textwidth}
        \centering
            \includegraphics[trim=4cm 8cm 4cm 8cm,clip,width=\linewidth]{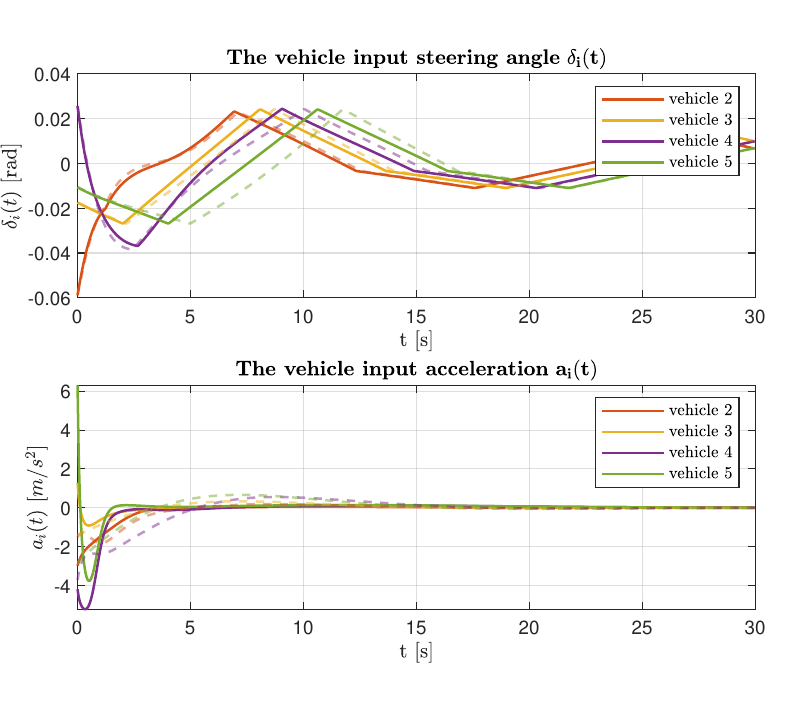}
        \caption{Scenario (A)}
        \label{fig:merg_input}
     \end{subfigure}
     \begin{subfigure}[b]{0.48\textwidth}
        \centering
            \includegraphics[trim=4cm 8cm 4cm 8cm,clip,width=\linewidth]{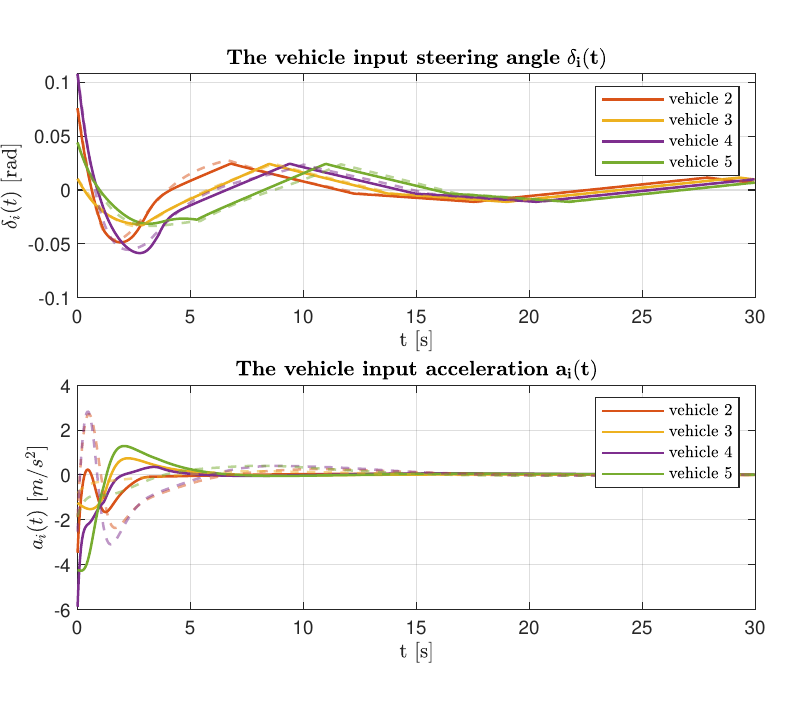}
        \caption{Scenario (B)}
        \label{fig:form_input}
     \end{subfigure}
    \caption{The vehicle input acceleration $a_i(t)$ and steering angle $\delta_i(t)$ for all $i = \{2,3,4,5\}$. The solid lines indicate results under the proposed controller \eqref{eq.control_lat}, \eqref{eq.control_lat_n}, \eqref{eq.control_lat_o}, \eqref{eq.lon_control_vir}, \eqref{eq.lon_control_vir_n}, and \eqref{eq.lon_control_vir_o}. The transparent dashed lines indicate the result under the baseline controller \eqref{eq.control_lat_n} and \eqref{eq.lon_control_vir_n}.}
        \label{fig:input_res}
\end{figure*}


\section{Experimental Results}\label{sec.experiment}
This section presents the experimental validation of the proposed methods using connected vehicles. The control algorithms are designed for full-scale autonomous cars with acceleration inputs. However, due to limitations of the available testbed, the experiments were conducted indoors on a fleet of scale-model SVEA vehicles—1:10 scale miniature platforms with velocity control inputs. While the SVEA platforms differ from full-scale cars, they are sufficient to demonstrate the effectiveness of the proposed methods as a proof of concept.

\begin{figure}[t]
	\centering	\centerline{\includegraphics[trim={4cm 5cm 6cm 3cm},clip,width=1\linewidth]{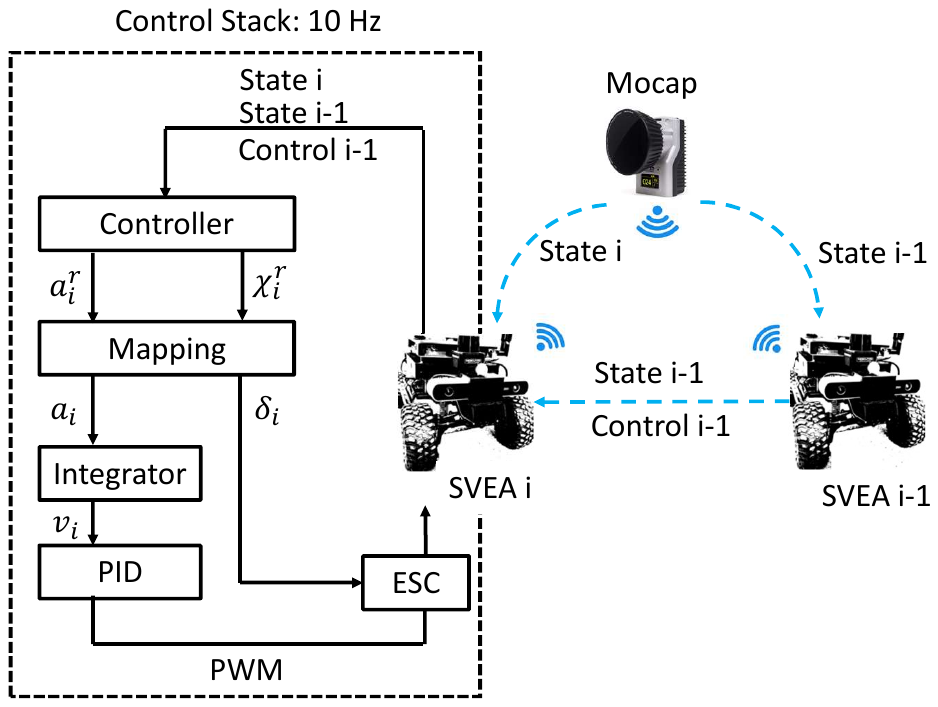}}
	\caption{Control and communication layout of the experimental studies.}
	\label{fig:exp_control_layout}
\end{figure}

Each SVEA is equipped with a Zed Box powered by an NVIDIA Jetson-embedded computer where the proposed controllers run onboard. Qualisys motion capture system (MOCAP) is used to provide accurate real-time state estimation $(\bold p_i, v_i, \theta_i)$ to each vehicle through communication. Each vehicle $i$ shares its state information $(\bold p_i, v_i)$ to its neighbor vehicle $i-1$ according to the communication topology defined in Assumption \ref{fig:topology}. Both the MOCAP-to-vehicle and vehicle-to-vehicle communication are enabled through a local WiFi network using the ROS framework and a tailored communication protocol. The control algorithms $\chi_i$ and $a_i$ are implemented in Python onboard each SVEA with an update frequency fixed at 10 Hz. Since the DC motor on the SVEA platform receives pulse-width modulation (PWM) signals from the electronic speed controller (ESC) to control its shaft speed, the input acceleration $a_i$ is integrated and processed to generate the corresponding PWM signal for accurate speed tracking. The experimental study's overall control and communication stack is shown in Fig. \ref{fig:exp_control_layout}.

 \begin{figure}[t]
	\centering	\centerline{\includegraphics[trim={7cm 3cm 6cm 3cm},clip,width=1\linewidth]{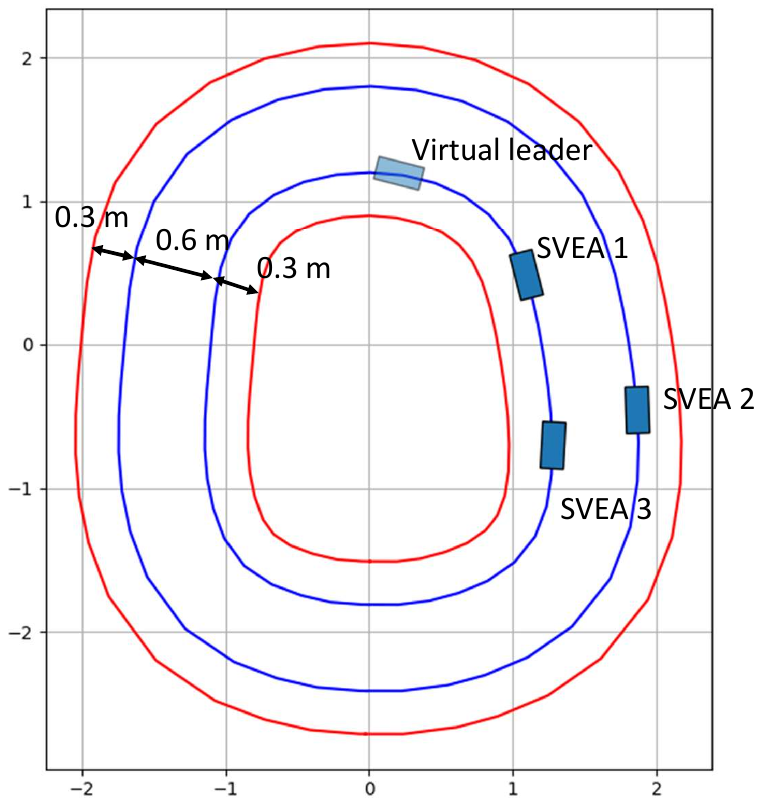}}
	\caption{The platoon merging scenario of 3 SVEAs for the experimental studies.}
	\label{fig:exp_scenario}
\end{figure}


The layout of the experiment is shown in Fig. \ref{fig:exp_scenario}. Due to the limitation of lab space, we designed the layout to be a circular-shaped path for continuous driving and testing. As shown in Fig. \ref{fig:exp_scenario}, the red lines indicate the road boundaries, while the blue lines are the reference paths. The reference paths are placed 0.6 \si{\ meter} apart and 0.3 \si{\ meter} away from the left respective right road edges. The vehicles are configured to resemble a platoon merging scenario. Here, SVEA 1 and SVEA 3 are designed to drive in a platoon on the inner reference path following a virtual platoon leader simulated onboard SVEA 1, which travels at constant velocity 0.7 \si{\metre\per\second}. SVEA 2 is initially placed on the outer reference path with a higher velocity and is designed to merge between the two vehicles given a merging command. The merging command is activated when the longitudinal distance between SVEA 1 and SVEA 2 is less than \SI{0.8}{\cm}. 

The controller gains tuned for the experiment scenario are $k_1 = 5$, $k_2 = 2$, $k_3 = 0.5$, $k_4 = 1$, $k_5 = 3$, $k_6 = 0.2$. The desired velocity is $v^*=v_i^*=\SI{0.7}{\metre\per\second}$ and the desired relative arc length $e_i^*=\SI{1.5}{\metre}$. The wheel base for each vehicle is $L_i = \SI{0.324}{\metre}$ and the safe margin is chosen as $\epsilon = \SI{0.6}{\metre}$, $\epsilon_w = \SI{0.15}{\metre}$. The initial velocity of SVEA 2 before merging the platoon is $v_2(0)=\SI{1.2}{\metre\per\second}$.

\begin{figure*}[t]
     \centering
     \begin{subfigure}[b]{0.48\textwidth}
        \centering
            \includegraphics[trim=4cm 4.5cm 4cm 4cm,clip,width=\linewidth]{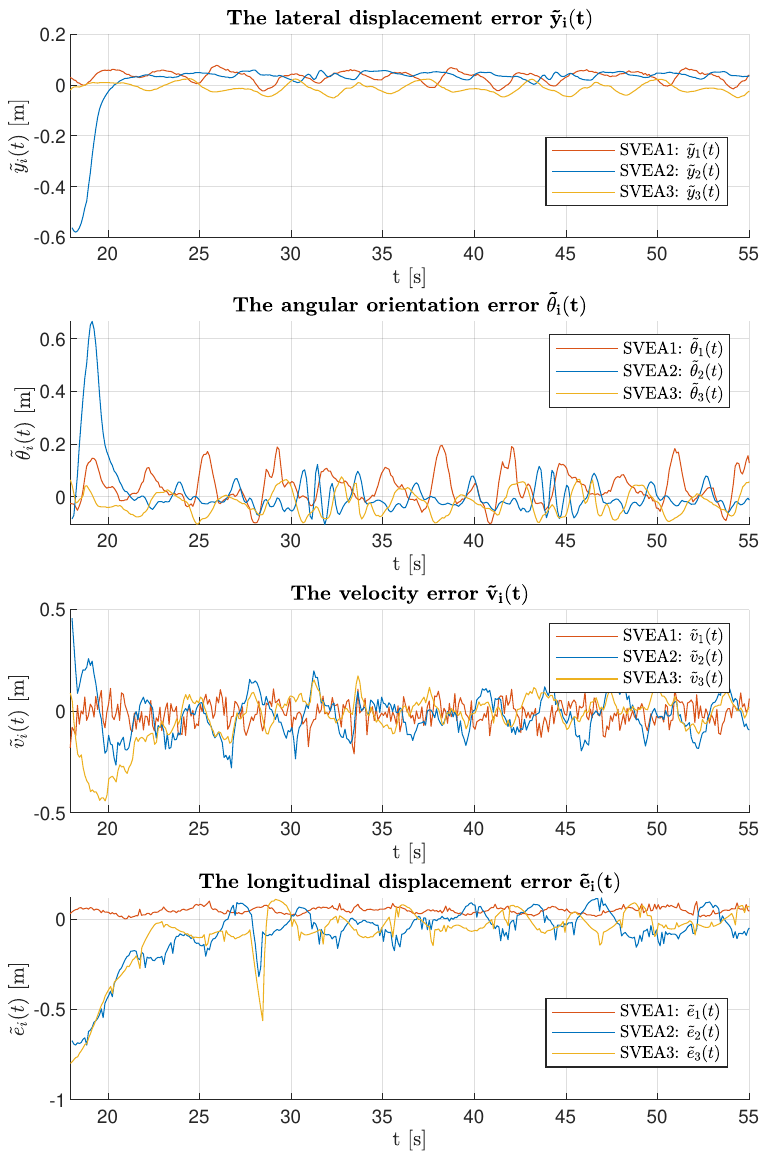}
        \caption{Experiment with the safe platoon controller}
        \label{fig:error_safe_exp}
     \end{subfigure}
     \begin{subfigure}[b]{0.48\textwidth}
        \centering
            \includegraphics[trim=4cm 4.5cm 4cm 4cm,clip,width=\linewidth]{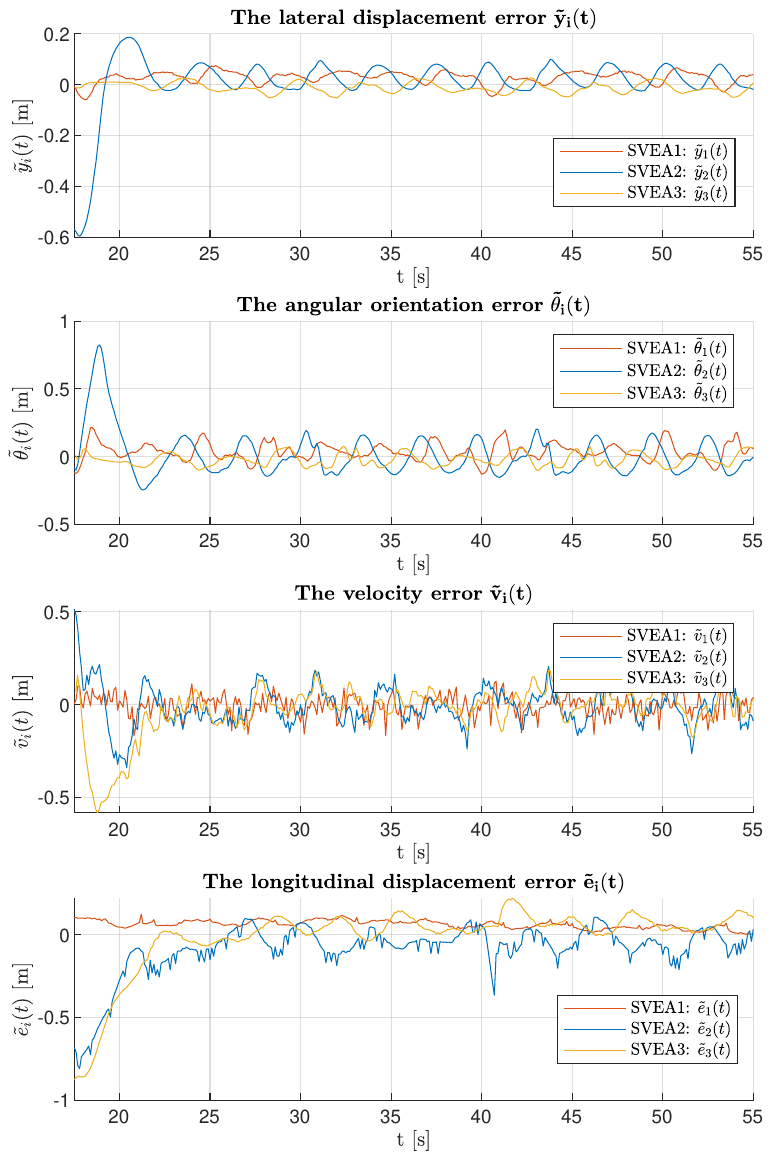}
        \caption{Experiment with the baseline controller}
        \label{fig:error_nom_exp}
     \end{subfigure}
    \caption{Time evolution of the lateral position error $\tilde{y}_i(t)$, the angular error $\tilde{\theta}_i(t)$, the velocity error $\tilde{v}_i(t)$, and the relative formation error $\tilde{e}_i(t)$ for SVEA vehicles in the experiment.}
        \label{fig:error_res_exp}
\end{figure*}

\begin{figure*}[t]
     \centering
     \begin{subfigure}[b]{0.48\textwidth}
        \centering
            \includegraphics[trim=4cm 8.5cm 4cm 8.5cm,clip,width=\linewidth]{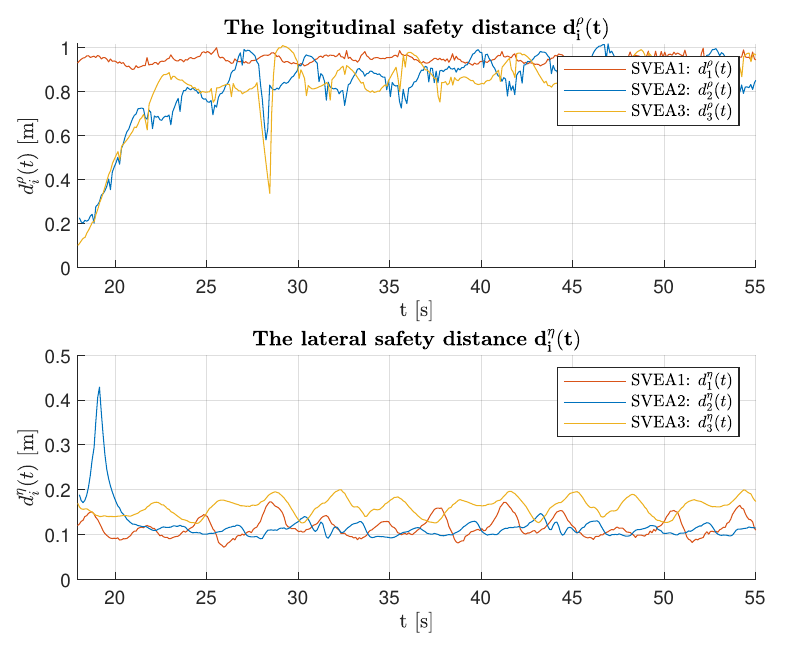}
        \caption{Experiment with the safe platoon controller}
        \label{fig:dist_safe_exp}
     \end{subfigure}
     \begin{subfigure}[b]{0.48\textwidth}
        \centering
            \includegraphics[trim=4cm 8.5cm 4cm 8.5cm,clip,width=\linewidth]{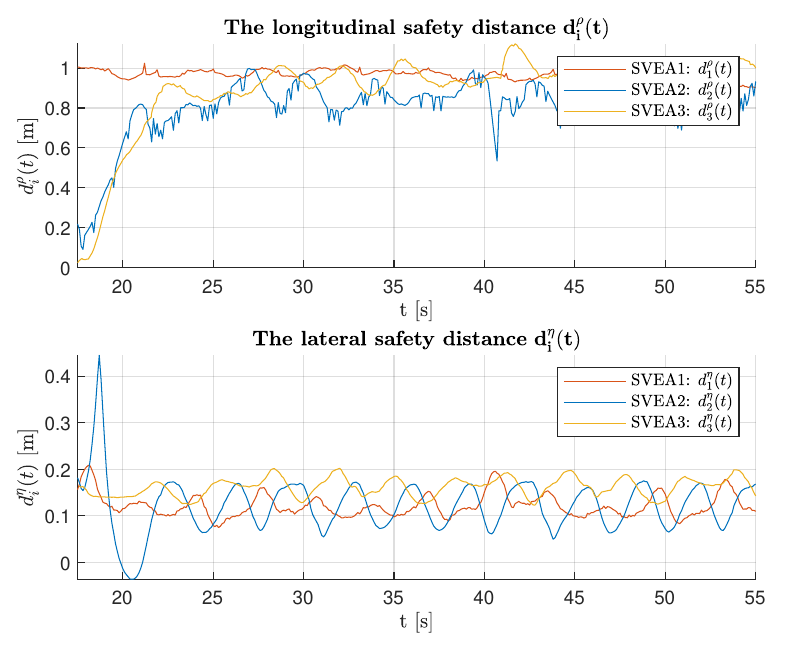}
        \caption{Experiment with the baseline controller}
        \label{fig:dist_nom_exp}
     \end{subfigure}
    \caption{Time evolution of the safety distance $d^{\rho}_i(t)$ and $d^\eta_i(t)$ for SVEA vehicles in the experiment.}
        \label{fig:dist_res_exp}
\end{figure*}

\begin{figure*}[t]
     \centering
     \begin{subfigure}[b]{0.48\textwidth}
        \centering
            \includegraphics[trim=3.5cm 8.5cm 4cm 8cm,clip,width=\linewidth]{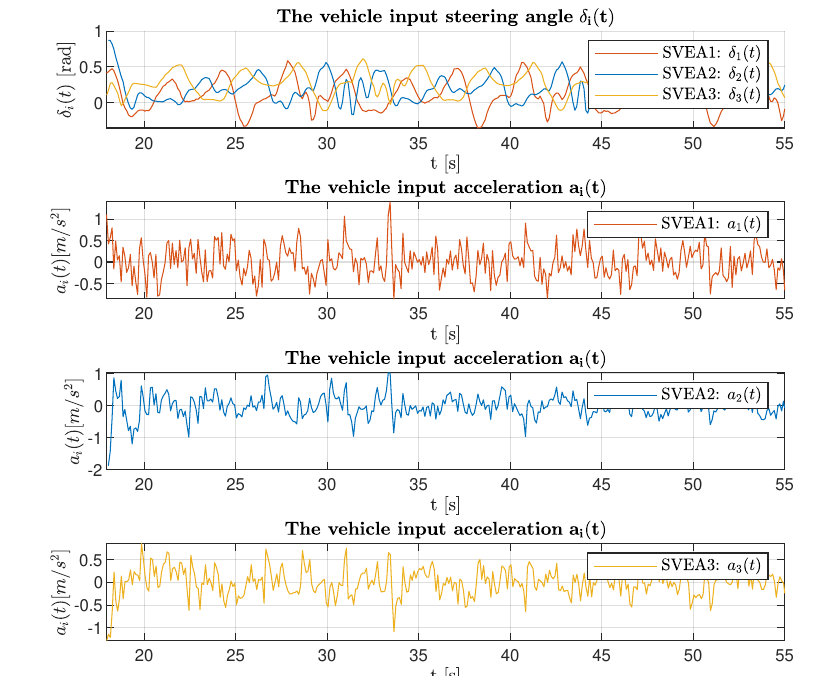}
        \caption{Experiment with the safe platoon controller}
        \label{fig:exp_input_safe}
     \end{subfigure}
     \begin{subfigure}[b]{0.48\textwidth}
        \centering
            \includegraphics[trim=3.5cm 8.5cm 4cm 8cm,clip,width=\linewidth]{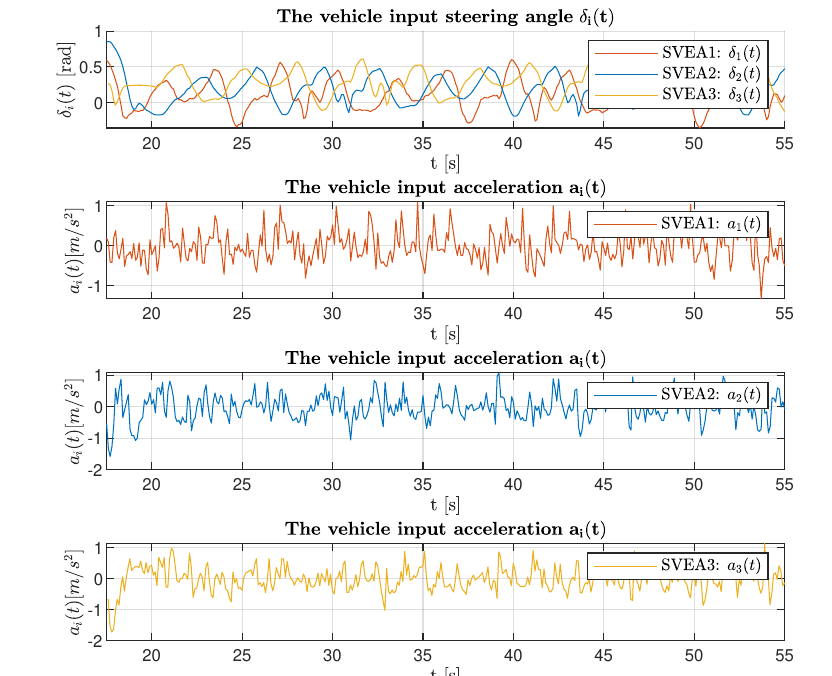}
        \caption{Experiment with the baseline controller}
        \label{fig:exp_input_nom}
     \end{subfigure}
    \caption{The vehicle input acceleration $a_i(t)$ and steering angle $\delta_i(t)$ for SVEA vehicles in the experiment. The input acceleration $a_i(t)$ is displayed in separate plots for clarity. }
        \label{fig:exp_input}
\end{figure*}

To analyze the effectiveness of the proposed method, we again compare the proposed safe platoon controller with the nominal controller as the baseline. The experimental results are presented from the time when the merging command is activated. 
As shown in Fig. \ref{fig:error_res_exp}, SVEA vehicles' states converge to desired values under both the safe platoon controller and baseline controller.
It demonstrates the robustness of the proposed controller under inaccurate actuation of the miniature vehicle. For safety, in Fig. \ref{fig:dist_safe_exp}, the proposed controller guarantees both $d^\rho_2$ and $d^\eta_2$ positive, resulting in a safe merging process. In comparison, the baseline controller failed to keep $d^\eta_2$ above zero as shown in Fig. \ref{fig:dist_nom_exp}, resulting in SVEA 2 driving over the road boundary during merging. For longitudinal safety distance $d^\rho_i$ between vehicles, due to safety, we did not tune the experimental setting to stress test the difference in terms of inter-vehicle safety between the proposed controller and the baseline controller. But the smaller value of $d^\rho_i$ in Fig. \ref{fig:dist_nom_exp} indicates a higher collision risk of the baseline controller. Finally, the input acceleration and steering are shown in Fig. \ref{fig:exp_input}. The proposed safe platoon controller and the baseline generate similar input signals with reasonable magnitude.

The experiments are recorded and the video can be found at \url{https://bit.ly/platoon_formation_experiment}. In summary, we conclude that the proposed safe platoon controller is applicable in practical implementations.

\section{Conclusion}\label{sec.conclusion}
This study addresses the problem of platoon formation on multi-lane roads with a general curved shape. The controller design is structured into two key components: path-following and longitudinal formation. To ensure safety, a constructive barrier feedback mechanism is introduced for both components, enabling lateral avoidance of road boundaries and longitudinal avoidance of preceding vehicles. Theoretical and numerical analyses are conducted to evaluate the performance and safety properties of the proposed method. Additionally, the method is experimentally validated using miniature vehicles, demonstrating its practical applicability in real-world scenarios.

Future research will focus on enhancing the controller’s capabilities by incorporating on-road obstacle avoidance during formation. Furthermore, a deeper analysis of the method’s robustness against model uncertainties and measurement errors is essential to further strengthen its reliability in practical implementations. 

\appendix
\renewcommand{\thesubsection}{\Alph{subsection}}
\subsection{Proof of Lemma \ref{lem.boundedv}}\label{app.1}
\begin{proof}
Recall \eqref{eq.dyna_frenet_lon_ref}, \eqref{eq.lon_control_vir}, \eqref{eq.lon_control_vir_n}, and \eqref{eq.lon_control_vir_o}, the closed-loop dynamics of the state $(\tilde e_i,\nu_i)$ is expressed as
\begin{equation}\label{eq.ei.nui}
    \left\{
    \begin{aligned}
        &\dot{\tilde e}_i = \nu_i\\
        &\dot{\nu}_i = -k_4\tilde e_i-k_5\nu_i-k_6\frac{\dot d_i^{\rho}}{d_i^{\rho}}
    \end{aligned}
    \right.  
\end{equation}
Consider the following Lyapunov function
    \begin{equation}\label{eq.Llong}
        \mathcal{L}_i = \frac{1}{2}k_4\Tilde{e}_i^2+\frac{1}{2}\nu_i^2
    \end{equation}
Recall \eqref{eq.ei.nui} and use the fact that $\dot{d}_i^{\rho} = \nu_i$, one has
\begin{equation}
    \dot{\mathcal{L}}_i = -k_5\nu_i^2 - k_6\frac{\nu_i^2}{d_i^{\rho}}
\end{equation}
which is negative semi-definite as long as $d_i^{\rho}>0$. Hence one concludes that $(\tilde e_i,\nu_i)$ is bounded $\forall t\ge 0$ provided $d_i^{\rho}>0$ for all $t>0$.

Proof of item (1):\\
Analyze the derivatives of $\dot d_i^{\rho}=\nu_i$, one has
\begin{equation}\label{eq.ddot_dg}
    \ddot{d}_i^{\rho} = -k_6\frac{\dot{d}_i^{\rho}}{d_i^{\rho}}-k_4\Tilde{e}_i-k_5\nu_i
\end{equation}
We will prove that $d_i^{\rho}$ remains positive using proof by contradiction. 
Assume there is a finite time $T>0$ such that $d_i^{\rho}(T)$ approaches zero. Take the integral of \eqref{eq.ddot_dg} from 0 to $T$, we get
\begin{equation}
\begin{aligned}
     k_6(\ln d_i^{\rho}(T)-\ln d_i^{\rho}(0)) = & \dot{d}_i^{\rho}(0) - \dot{d}_i^{\rho}(T)\\
     &-\int_0^T(k_4\Tilde{e}_i+k_5\nu_i)d\tau
\end{aligned}
\end{equation}
The left-hand side of the above equation tends to negative infinity, while the right-hand side is either bounded or tends to positive infinity since $\Tilde{e}_i$ and $\nu_i$ are bounded for all $0<t<T$, and $\dot{d}_i^{\rho}$ is either bounded or negative infinity. This leads to a contradiction, hence, $d_i^{\rho}$ remains positive for all time. Therefore, one concludes that $(\tilde e_i,\nu_i)$ are bounded. From there, direct application of \cite[Lemma 2]{Zhiqi2023Constructive}, one ensures that  $\frac{\dot{d}_i^{\rho}}{d_i^{\rho}}$ remains bounded for all the time. This in turn, implies that $a_i^r$ \eqref{eq.lon_control_vir} is also bounded for all the time.

Proof of item (2):

Using a similar argument as the proof  of item (2)-\cite[Lemma 2]{Zhiqi2023Constructive} along with Assumption \ref{ass.vehicle_index},   one concludes that the unique equilibrium point $(\tilde{e}_i, \nu_i)=(0,0)$ is asymptotically stable. 

\end{proof}


\subsection{Proof of theorem 1}\label{app.2}

\begin{proof}
Consider first the lateral error state $(\Tilde{y}_i, \Tilde{\theta}_i)$, by using \eqref{eq.dyna_frenet},\eqref{eq.control_lat}, \eqref{eq.control_lat_n}, and \eqref{eq.control_lat_o}, its closed-loop dynamics can be expressed as
\begin{equation}\label{eq.closed_lat_error}
    \left\{
    \begin{aligned}
        \dot{\Tilde{y}}_i &= v_i\sin\Tilde{\theta}_i\\
        \dot{\Tilde{\theta}}_i &= -k_1v_i\Tilde{y}_i\frac{\sin\Tilde{\theta}_i}{\Tilde{\theta}_i} - k_2|v_i|\Tilde{\theta}_i\\
        &-k_3 |v_i|\left (\frac 1 {d^{\eta_L}_i}+\frac 1 {d^{\eta_R}_i}\right)\sin\tilde \theta_i
    \end{aligned}
    \right.
\end{equation}
Define the following Lyapunov function:
 \begin{equation}\label{eq.Llat}
    \mathcal{L} = \frac{1}{2}k_1\Tilde{y}_i^2+\frac{1}{2}\Tilde{\theta}_i^2
\end{equation}
Together with \eqref{eq.closed_lat_error}, the derivative of \eqref{eq.Llat} is given by: 
\begin{equation}\label{eq:dotL}
    \dot{\mathcal{L}} = -k_2 |v_i|\Tilde{\theta}_i^2-k_3 |v_i|\left (\frac 1 {d^{\eta_L}_i}+\frac 1 {d^{\eta_R}_i}\right )\Tilde{\theta}_i\sin\Tilde{\theta}_i
\end{equation}
which is negative semi-definite given that $d^{\eta_R}_i>0$, $d^{\eta_R}_i>0$, and $|\Tilde{\theta}_i|\leq \pi/2-\epsilon_1$ (and even more so when $|\Tilde{\theta}_i|\leq \pi$). One verifies that the state $(\tilde y_i,\tilde \theta_i)$ is bounded as long as $d^{\eta_R}_i>0$, and $d^{\eta_R}_i>0$. 

Proof of item (1):\\
Since Lemma \ref{lem.boundedv}- item 1 concludes that $d_i^{\rho}$ remains positive all the time. We will focus on showing the bounds of $|\tilde \theta_i|$ and $|\tilde y_i|$ as well as the positiveness of $d^{\eta_L}_i$ and $d^{\eta_R}_i$. Since the Lyapunov function \eqref{eq.Llat} is nonincreasing,  the initial condition $k_1 \tilde y_i^2(0)+\tilde \theta_i^2(0)<(\frac{\pi}{2}-\epsilon_1)^2$ implies that $|\tilde \theta_i(t)|<\pi/2-\epsilon_1$ as long $d^{\eta_L}_i>0$ and $d^{\eta_R}_i>0$. Recall that from \eqref{eq.deta} and Assumption \ref{ass.road_shape}, if $d^{\eta_L}_i>0$ and $d^{\eta_R}_i>0$ one ensures that $|\Tilde{y}_i|< \frac{1}{\chi_i^r}$.

Due to the convex combination $d^{\eta_R}_i+d^{\eta_L}_i=2w-2\epsilon_w$ and the symmetry of the problem at hand, proving $d^{\eta_R}_i>0$ is similar to prove $d^{\eta_L}_i>0$. From now on, we will only show the proof for $d^{\eta_L}_i$ and denote $d^{\eta}_i=d^{\eta_L}_i$ for the sake of simplicity. Take the derivative of ${d^{\eta}_i}^{'} =- \sign(v_i)\sin\Tilde{\theta}_i$ with respect to $\sigma$ to obtain\footnote{The derivative of $\sign(v_i)$ with respect to $\sigma$ is set to zero because when $v_i=0$ the system is at rest and $\sigma$ cannot be used as a substitute for time to describe changes.}:
\begin{equation}
\begin{aligned}
      {d^{\eta}_i}^{''} =&- \sign(v_i)\frac d {d\sigma}\sin \tilde\theta_i
\end{aligned}
\end{equation}
since $\frac d {d\sigma}\sin \tilde\theta_i=\cos\tilde \theta \sign(v_i)\left(\chi_i - \frac{\chi_i^r\cos\Tilde{\theta}_i}{1-\chi_i^r\Tilde{y}_i}\right)$
\begin{equation}\label{eq.ddot_deta}
\begin{aligned}
      {d^{\eta}_i}^{''}
      = &-k_3\cos\Tilde{\theta}_i\frac{{d^{\eta}_i}^{'}}{d^{\eta}_i}-\alpha_i
\end{aligned}
\end{equation}
where 
\begin{align*}
\alpha_i(\sigma(t))=&-\cos\Tilde{\theta}_i ( -k_1\frac{\sin\Tilde{\theta}_i}{\Tilde{\theta}_i}\Tilde{y}_i- k_2\sign(v_i)\Tilde{\theta}_i\\
&-k_3\frac{\sign(v_i)\sin\tilde \theta_i}{d_i^{\eta_R}})
\end{align*}

Take integral of the above equation with respect to $\sigma$ from $\sigma_0=\sigma(0)$ to $\sigma_T=\sigma(T)$ with a finite $\sigma_T$ (or equivalently finite $T$), one has:
 \begin{align*}
    & k_3\int_{\sigma_0}^{\sigma_T}\cos\Tilde{\theta}_i\frac{{d^{\eta}_i}^{'}}{d^{\eta}_i} d \sigma={d^{\eta}_i}^{'}(T)-{d^{\eta}_i}^{'}(0)-\int_{\sigma_0}^{\sigma_T}\alpha_id\sigma
    \end{align*}
Using the fact that $|\Tilde{\theta}|<\frac{\pi}{2}$ in the interval $[\sigma_0,\sigma_T)$, and $\cos\Tilde{\theta} >0$ with a bounded derive, one ensures that there exists $\bar k_3\in (0,k_3]$ a positive and bounded scalar such that $\int_{\sigma_0}^{\sigma_T}k_3\cos\Tilde{\theta}_i\frac{{d^{\eta}_i}^{'}}{d^{\eta}_i}d\sigma=\bar k_3\ln \frac{d^{\eta}_i(\sigma_T)}{d^{\eta}_i(\sigma_0)}$ and hence: 
\begin{equation}\label{eq.detaint}
    \begin{aligned}
    &\bar k_3\ln \frac{d^{\eta}_i(\sigma_T)}{d^{\eta}_i(\sigma_0)}={d^{\eta}_i}^{'}(\sigma_T)-{d^{\eta}_i}^{'}(\sigma_0)-\int_{\sigma_0}^{\sigma_T}\alpha_id\sigma
    \end{aligned}
\end{equation}
From this and using the fact that $\frac{{d^{\eta}_i}^{'}(\sigma_0)}{d^{\eta}_i(\sigma_0)}$, ${d^{\eta}_i}^{'}(\sigma_T)$, and ${d^{\eta}_i}^{'}(\sigma_0)$ are bounded and that $\alpha_i$ is bounded in  $[\sigma_0,\sigma_T)$, one concludes that $d^{\eta}_i$ remains positive by following the same reasoning as in the proof of Lemma 1, Item 1.

Proof of item (2):\\
Since $\chi_i^n$ is bounded because $\tilde y_i$ and $\tilde \theta_i$ are bounded, to show that controller $\chi_i$ is bounded, it suffices to show $\chi_i^c$ is bounded when $d_i^{\eta}$ converges to zero as $\sigma(t)$ tends to infinity. The Proof follows a similar structure of \cite[Lemma 2]{Zhiqi2023Constructive}. We will first show that ${d_i^{\eta}}^{'}$ converges to zero.
Define a new variable $\mu$ such that $\mu=\int_{\sigma_0}^{\sigma_t}\frac{1}{d^{\eta}_i}d \sigma$ and rewrite \eqref{eq.ddot_deta} as
\begin{equation}
\begin{aligned}
      \frac{d}{d\mu}{d^{\eta}_i}^{'} 
      = &-k_3\cos\Tilde{\theta}_i{d^{\eta}_i}^{'}-o(\mu)
\end{aligned}
\end{equation}
with $k_3\cos\Tilde{\theta}_i>0$ and $o(\mu)=d_i^\eta \alpha_i$ a perturbation term that tends to zero whens $\mu$ goes to infinity (or equivalently when $d_i^{\eta}$ converges to zero). One recognizes that this is the dynamics of a first-order system (with ${d^{\eta}_i}^{'}$ the state) perturbed by a vanishing perturbation. From there, one concludes that ${d^{\eta}_i}^{'}$ is bounded and also converges to zero as $d_i^{\eta}$ converges to zero. To prove that is $\chi_i^c$ is bounded when $({d^{\eta}_i}^{'}, d^{\eta}_i)\to(0,0)$, it suffices to show that $\frac{{d^{\eta}_i}^{'}}{d^{\eta}_i}$ is bounded. Hence, differentiating $\frac{{d^{\eta}_i}^{'}}{d^{\eta}_i}$ with respect to $\mu$, one verifies that:
\begin{equation}
\begin{aligned}
      \frac{d}{d\mu}\frac{{d^{\eta}_i}^{'}}{d^{\eta}_i} = -(k_3\cos\Tilde{\theta}_i+ {d^{\eta}_i}^{'})\frac{{d^{\eta}_i}^{'}}{d^{\eta}_i}-\alpha_i
\end{aligned}
\end{equation}
Using the fact that $\alpha_i$ is bounded and ${d^{\eta}_i}^{'}$ is converging to zero, one ensures that there exists a $\mu^*$ or equivalently a $\sigma^*$ (and therefore a time instant $T$), such that $({d^{\eta}_i}^{'} +k_3 \cos \Tilde\theta) >0, \; \forall \mu \geq \mu^*$. From there, one guarantees that $\frac{{d^{\eta}_i}^{'}}{d^{\eta}_i}$  is bounded, hence, $\chi_i$ is bounded and well-defined for all the time.

To prove $a_i$ is bounded and well-defined, we recall equation \eqref{eq.ar2a}, \eqref{eq.vi}, and \eqref{eq.dyna_frenet}. One concludes that $a_i$ is well-defined and bounded since $|\Tilde{\theta}_i|<\frac{\pi}{2}-\epsilon_1$, $|\Tilde{y}_i|<\frac{1}{\chi_i^r}$, and $a_i^r$ and $v_i^r$ are bounded and well-defined (Lemma \ref{lem.boundedv}).

Proof of item (3): 

Using Barbalat’s Lemma along with a similar argument as the proof of Lemma 2 - item (2) in \cite{Zhiqi2023Constructive}, one concludes that the unique equilibrium point $(\tilde{y}_i, \Tilde{\theta}_i)=(0,0)$  for the lateral subsystem is asymptotically stable. 


From Lemma \ref{lem.boundedv} item (2), one concludes that the origin of $(\tilde e_i,\nu_i)$ is asymptotically stable. Recall equation \eqref{eq.vr}, one has $\tilde v_i^r=\tilde v_i$ provided $(\tilde{y}_i, \Tilde{\theta}_i)=(0,0)$. From Assumption \ref{ass.platoon_form}, the leader vehicle is already in its desired states, i.e., $\tilde s_1=0$ and $\tilde v_1=\tilde v_1^r=0$, hence, the asymptotic stability of  $(\tilde e_i,\nu_i)=(\tilde s_{i-1}-\tilde s_i,\tilde v_{i-1}^r-\tilde v_i^r)=(0,0)$ and $(\tilde{y}_i, \Tilde{\theta}_i)=(0,0)$  implies that the origin of $(\tilde s_i,\tilde v_i)$ is also asymptotically stable.

\end{proof}

\addtolength{\textheight}{0cm}   





\end{document}